%% file: cas-dc-template.tex
\documentclass[a4paper,fleqn]{cas-sc}

\usepackage[authoryear,longnamesfirst]{natbib}
\usepackage{standalone}
\usepackage{subcaption}
\usepackage{bm}
\usepackage{siunitx}
\usepackage{cleveref}
\usepackage{graphicx}
\graphicspath{{./figures/}}
\usepackage{amsmath}
\usepackage{bigints}
\usepackage{mathtools}
\usepackage{pgfplots}
\usepackage{bm}
\usepackage{xargs} 
\usepackage{soul}
\usepackage{widetext}
\usepackage{tikz}
\usetikzlibrary{shapes,arrows}
\usepackage{multicol}
\usepackage{cuted}
\pgfplotsset{compat=newest}
\usepgfplotslibrary{colorbrewer}
\usepgfplotslibrary{groupplots}
\def\tsc#1{\csdef{#1}{\textsc{\lowercase{#1}}\xspace}}
\tsc{WGM}
\tsc{QE}
\tsc{EP}
\tsc{PMS}
\tsc{BEC}
\tsc{DE}

\begin{document}
\let\WriteBookmarks\relax
\def\floatpagepagefraction{1}
\def\textpagefraction{.001}

\shorttitle{RTF and noise modeling of IPUTs}

\shortauthors{S V Valappil et~al.}

\title [mode = title]{Semi-analytical modeling of receive transfer function and thermal noise of integrated photonic ultrasound transducers}                      



%
\author[1]{Sabiju Valiya Valappil}[type=author,
                        auid=000,bioid=1,
                        orcid=0000-0001-7511-2910]

\cormark[1]

\fnmark[1]

\ead{S.ValiyaValappil@tudelft.nl}
\credit{Conceptualization of this study, Methodology, Software, Writing - Original draft preparation}

\affiliation[1]{organization={Faculty of Applied Sciences, Delft University of Technology},
    addressline={Lorentzweg 1}, 
    city={Delft},
    postcode={2628 CJ}, 
    country={The Netherlands}}

\author[2]{Peter Harmsma}
\fnmark[2]
\ead{peter.harmsma@tno.nl}
\credit{Funding acquisition, Conceptualization, Supervision, Writing - Review \& Editing}
\affiliation[2]{organization={Nederlandse Organisatie voor Toegepast Natuurwetenschappelijk Onderzoek (TNO)},
    addressline={Stieltjesweg 1}, 
    city={Delft},
    postcode={2628CK}, 
    country={Netherlands}}

\author[3]{Maurits van der Heiden}
\fnmark[3]
\ead{maurits.vanderheiden@tno.nl}
\credit{Funding acquisition, Conceptualization, Supervision, Writing - Review \& Editing}

\author[1]{Martin Verweij}
\fnmark[4]
\ead{M.D.Verweij@tudelft.nl}
\credit{Funding acquisition, Conceptualization, Supervision, Writing - Review \& Editing}

\author[3]
{Paul van Neer}
\fnmark[5]
\ead{paul.vanneer@tno.nl}
\credit{Funding acquisition, Conceptualization, Supervision, Writing - Review \& Editing}
\affiliation[3]{organization={Nederlandse Organisatie voor Toegepast Natuurwetenschappelijk Onderzoek (TNO)},
    addressline={Oude Waalsdorperweg 63}, 
    city={Den Haag},
    postcode={2509JG}, 
    country={Netherlands}}

\cortext[cor1]{Corresponding author}




\begin{abstract}
Ultrasound transducers (UTs) are extensively used in several applications across a multitude of disciplines. A new type of UTs namely integrated photonic ultrasound transducers (IPUTs) possess superior performance due to the presence of optical interrogation systems, avoiding electric crosstalk and thermal electronic noise of the sensor. However, a major component of the IPUT's noise floor is its thermal acoustic noise. Several studies have been proposed to characterize IPUTs' behavior; nevertheless, these are either incomplete (model only the thermal noise) or targeted to characterize specific responses such as static behavior, in which the modeled receive transfer function (RTF) is about two orders lower than the experiments. In this study, we develop semi-analytical models based on time-domain finite element analysis and analytical expressions to characterize the RTF and thermal noise-induced noise equivalent pressure of IPUTs. We validate the models by comparing them with the literature, where we obtain a close match between them.
\end{abstract}




\maketitle

\section{Introduction}

Ultrasound sensors are extensively used in various applications, such as medical diagnostics~\cite{Carovac2011-sz}, non-destructive evaluation~\cite{DRINKWATER2006525}, flow measurement~\cite{SANDERSON2002125}, distance measurement~\cite{Carullo2001AnUS}, and food processing~\cite{AWAD2012410}, among others. Since these sensors do not apply ionizing radiation, are often real-time, and are relatively inexpensive, ultrasonic imaging is used as a common medical imaging modality. The global market size for diagnostic medical ultrasound was US\$6.7 billion in 2020 and is expected to grow to US\$9.3 billion in 2027~\cite{Insight_2022,Markets_2023}. Ultrasound sources used in medical imaging have limited transmitted signal strength due to the safety limit. Additionally, the attenuation rate inside the human tissue is high, which drastically decreases the signal strength during deep-tissue imaging. Thus, the receivers need to possess a high signal-to-noise ratio (SNR), for which a significant transfer function (the output response due to the incoming pressure) and low noise floor are required. 

Conventional ultrasound transducers are based on piezoelectric systems that convert acoustic pulses to electrical signals and vice versa~\cite{Tressler1998}. Additionally, micro-electronic mechanical systems (MEMS)-based transducers are also being used for medical imaging. These transducers are divided into capacitive micromachined ultrasonic transducers (CMUTs)~\cite{Ilkhechi:20} and piezoelectric micromachined ultrasonic transducers (PMUTs)~\cite{Jung_2017} depending on the physical signal conversion principle. CMUTs rely on capacitive changes while PMUTs use thin-film piezoelectric layers to convert the membrane deformation to electrical pulses. The noise equivalent pressure (NEP)---the pressure value for which the SNR is unity---for the aforementioned devices operating around \SI{1}{\mega\hertz} with \SI{80}{\percent} \SI{-6}{\decibel} bandwidth (BW) is comparable to the standard piezoelectric-based ultrasound transducers, which is around \SI{0.5}{\pascal}~\cite{Ilkhechi:20,https://doi.org/10.1118/1.4792462,Manwar2020-vy}. Additionally, their transmit transfer functions are also in the same order, while their noise floor is limited by the thermal electric noise, which is much higher than their thermal acoustic noise (due to their limited size). A novel type of ultrasound transducers -- integrated photonic ultrasound transducers (IPUTs) may sidestep the traditional transducer design trade-offs and limitations as they are based on different physical principles. 

IPUTs consist of a mechanical resonator combined with an optical integrated photonic circuit. The transfer function of the optical circuit is modified by the incoming pressure wave, which excites the mechanical resonator. This change in the optical transfer function can be measured using optical read-out concepts, for example using a laser~\cite{Westerveld2021}. Depending on the IPUT design and the intended read-out chain, they are generally classified as interferometer-based, such as a Mach-Zehnder interferometer (MZI)~\cite{Ouyang:19}, and whispering gallery mode-based as in the case of a ring resonator (RR)~\cite{Leinders2015}. MZIs comprise a sensing optical waveguide (attached to the mechanical resonator) and a reference optical waveguide. An optical probe signal emitted by a laser is distributed over the sensing- and reference waveguides. The signals from both arms interfere in a combiner, the output signals of which depend on the phase difference between the two paths. When the incoming acoustic pressure impinges on the IPUT, the sample waveguide deforms more than the reference waveguide because the former is on top of the mechanical resonator at or close to resonance. The deformation induces a length change and an effective refractive index change, leading to a phase shift between the reference and sensing signals. This phase shift is detected as an intensity variation at the interferometer outputs, which can be used to measure the incoming pressure. On the other hand, RR-based IPUTs feature a ring-shaped optical waveguide directly connected to the mechanical resonator. This ring-shaped waveguide forms a high-Q factor optical resonator with a series of resonance wavelengths having equidistant spacing in terms of optical frequency. Again, an acoustical pressure wave deforms the IPUT mechanical resonator, affecting both the length and effective refractive index of the optical resonator. This results in a shift of the optical resonance wavelengths. By applying a laser on the steep flank of one such resonance, the aforementioned wavelength shift can be measured as an optical intensity change at the laser wavelength. The relation between the IPUT's measured quantity (e.g., optical phase shift) and the incident field (acoustic pressure) is described as its receive transfer function (RTF).

Several studies have been conducted to model the transfer function of IPUTs and similar optomechanical pressure sensors~\cite{Jansen2016MicroOptoMechanicalPS,7926285,9152628,Rochus_2018}. It is noteworthy that low-frequency opto-mechanical sensors are classified as pressure sensors. On the other hand, optomechanical sensors capable of detecting pressure pulses on or above ultrasound frequencies (i.e. above \SI{20}{\kilo\hertz}) are deemed as IPUTs. Jansen et al. proposed an MZI-based optomechanical pressure sensor with high measurement accuracy, where a theoretical model based on the static response of the membrane is used~\cite{Jansen2016MicroOptoMechanicalPS}. Since the static model will not be able to predict the dynamic behavior of the IPUTs, Zunic et al.~\cite{9152628} used an acousto-mechanical finite element model with analytical equations to design an IPUT consisting of a silicon nitride MZI placed on a silicon dioxide (SiO\textsubscript{2}) membrane for photoacoustic imaging. Rochus et al. modeled the optomechanical pressure sensors with the presence of residual stress using the finite element method (FEM)~\cite{7926285}. They also proposed an analytical procedure to design micro-opto-mechanical pressure sensors incorporating mechanical nonlinearities~\cite{Rochus_2018}. They further extended the MZI-based optomechanical devices to use as accelerometers where the modeling was carried out using FEM and analytical expressions~\cite{8724569}. Gao et al. developed equivalent circuit models for characterizing the behavior of optomechanical microphones that are also based on MZIs~\cite{8369932}. By following a similar modeling approach (FEM and analytical methods), Westerveld et al. studied the dynamic behavior of RR-based optomechanical devices with buckled acoustic membranes~\cite{8724528}. FEM and analytical modeling were further used to characterize other similar devices such as IPUTs with slotted RRs~\cite{ZHANG2017113,Lee2023} and Fabry-Perot resonators~\cite{Allen:11,Bitarafan:15}. However, all aforementioned analytical models either assume a static response of the membrane/plate or are based on the modal behavior of the membrane (considering the first mode). Additionally, all the FEM models listed here consider a harmonic response of the IPUT while also assuming that the membrane is the source, which in principle is not correct. In other words, instead of acoustic pulses arriving from the fluid medium to the IPUT (then radiating back to the fluid), the models assume that the membrane oscillates in the presence of fluid and emits acoustic signals to the fluid medium. Thus, to accurately characterize the IPUT's response, the model should capture a more realistic behavior.

To maximize the IPUT's sensitivity whilst maintaining a frequency bandwidth sufficient for medical imaging, we need to optimize its RTF. Next to this, it is desirable for a medical ultrasound system to have a low NEP. Hence it is important to understand, model, and predict the various noise contributions in an IPUT-based system. The main noise sources in an IPUT include the thermo-mechanical (otherwise called thermal) noise~\cite{PhysRevD.42.2437} and optical shot noise~\cite{RevModPhys.68.801}. The noise generated during the interrogation includes back action noise~\cite{doi:10.1126/science.1156032}, detector noise~\cite{doi:10.1063/1.1149175}, laser wavelength jitter noise~\cite{Keller:90}, and the laser relative intensity noise~\cite{refId0}. Additionally, depending on the environment of operation, the surrounding medium (e.g., air or water) can also induce thermal noise due to the Brownian motion of the molecules in the liquid or gas impinging the membrane~\cite{Davuluri:21}. The influence of the thermal noise on the sensitivity has been investigated extensively for various types of optomechanical devices because in strain, phase, and ultrasound measurement applications the thermal noise dominates the rest of the noise contributions~\cite{6994803,Hornig:22,yang2022micropascal}. On the contrary, in the case of optomechanical accelerometers, the optical shot noise dominates the other noises~\cite{Krause2012,7435565}. This is because these accelerometers operate at very high mechanical and optical Q-factors resulting in a reduced thermal noise. To accurately evaluate the contributions of these noises in the output signal of the IPUT, they need to be precisely characterized. Additionally, the thermal noise is intrinsic to the IPUT, i.e., it is related to the RTF. Thus, to predict the thermal noise correctly, we need to obtain accurate values for the IPUT's RTF as well. Moreover, to understand the influence of various parameter changes on different noise contributions and the RTF, we need to relate them to a parametric space. To that end, a detailed parametric model is needed. To the best of our knowledge, no accurate models exist that can characterize the actual wave propagation behavior of the IPUTs incorporating the acoustic radiation in the fluid medium. In addition, there are no models available that relate relevant IPUT parameters with the RTF, different noise types, and their interrelations. 

In this work, we present novel semi-analytical models to predict the RTF and noise contribution from the thermal noise of an IPUT in contact with a fluid (water) domain. The model is validated by comparing it to measurement results reported in the literature.
\section{Definition and operation of an IPUT sensor}
\label{sec:definition}
The majority of IPUTs integrate an RR or an MZI on their mechanical resonator. Often the mechanical resonator is a membrane. The modeling approach we employ here is generic to both cases; however, we focus on an RR-based IPUT design for clarity. An RR comprises an optical waveguide looped back to itself. Light from a straight waveguide is coupled to the RR when kept close due to the evanescent field effect. Therefore, an optical resonance with a high Q factor ($10^4$ to $10^6$) occurs when the resonator's optical path length is precisely a whole number of the wavelength. An incoming ultrasound pulse excites the membrane at (or close by) its mechanical resonance. The shift in the wavelength of the optical resonance as a function of the membrane deformation is then monitored. An optical detector placed at the output reads the optical intensity variation and provides the electrical measurements.

\begin{figure}[!ht]  
\centering
\pgfplotsset{cycle list/Dark2}
	\def\svgwidth{0.75\linewidth}
    \centering
	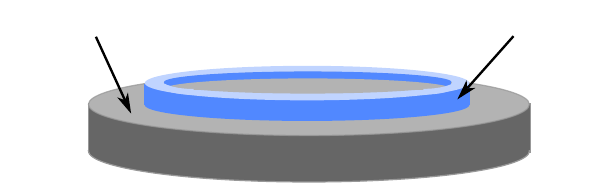

\caption{Schematic representation of an IPUT constituting an optical waveguide (blue ring) and a mechanical membrane (gray disc).}
\label{fig:iput}
\end{figure}
In this study, we define the IPUT, as shown in Figure~\ref{fig:iput}, which is composed of a mechanical resonator (membrane/disc) with an optical waveguide on top. We focus on modeling the deformation of the membrane and the induced changes in the transfer function of the optical waveguide located on top of the membrane. The optical transfer function is affected by the induced changes in length, shape, and refractive index of the waveguide~\cite{Wissmeyer2018}. As our IPUT only includes the membrane and optical waveguide, the noise contributions from the interrogation system, detectors, and other electronic components cannot be evaluated with this IPUT model. Additionally, the optical losses at the couplings from the optical fiber to the IPUT and the effects of the substrate-membrane boundary that can act as an elastic boundary are also neglected here for convenience. The operation of the IPUT is further elaborated next.

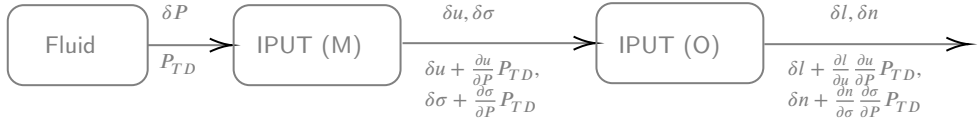
\begin{figure*}[!ht]
    \centering

\tikzset{every picture/.style={line width=0.75pt}} 

\begin{tikzpicture}[x=0.75pt,y=0.75pt,yscale=-1,xscale=1,scale=1]

\draw   (25,88.5) .. controls (25,84.08) and (28.58,80.5) .. (33,80.5) -- (87,80.5) .. controls (91.42,80.5) and (95,84.08) .. (95,88.5) -- (95,112.5) .. controls (95,116.92) and (91.42,120.5) .. (87,120.5) -- (33,120.5) .. controls (28.58,120.5) and (25,116.92) .. (25,112.5) -- cycle ;
\draw    (95,100.5) -- (135,100.5) ;
\draw [shift={(137,100.5)}, rotate = 180] [color={rgb, 255:red, 0; green, 0; blue, 0 }  ][line width=0.75]    (10.93,-3.29) .. controls (6.95,-1.4) and (3.31,-0.3) .. (0,0) .. controls (3.31,0.3) and (6.95,1.4) .. (10.93,3.29)   ;
\draw   (138,88.5) .. controls (138,84.08) and (141.58,80.5) .. (146,80.5) -- (213.5,80.5) .. controls (217.92,80.5) and (221.5,84.08) .. (221.5,88.5) -- (221.5,112.5) .. controls (221.5,116.92) and (217.92,120.5) .. (213.5,120.5) -- (146,120.5) .. controls (141.58,120.5) and (138,116.92) .. (138,112.5) -- cycle ;
\draw    (222.5,99.5) -- (317,99.5) ;
\draw [shift={(319,99.5)}, rotate = 180] [color={rgb, 255:red, 0; green, 0; blue, 0 }  ][line width=0.75]    (10.93,-3.29) .. controls (6.95,-1.4) and (3.31,-0.3) .. (0,0) .. controls (3.31,0.3) and (6.95,1.4) .. (10.93,3.29)   ;
\draw   (319.5,87.5) .. controls (319.5,83.08) and (323.08,79.5) .. (327.5,79.5) -- (395,79.5) .. controls (399.42,79.5) and (403,83.08) .. (403,87.5) -- (403,111.5) .. controls (403,115.92) and (399.42,119.5) .. (395,119.5) -- (327.5,119.5) .. controls (323.08,119.5) and (319.5,115.92) .. (319.5,111.5) -- cycle ;
\draw    (403.5,100) -- (503.5,100) ;
\draw [shift={(505.5,100)}, rotate = 180] [color={rgb, 255:red, 0; green, 0; blue, 0 }  ][line width=0.75]    (10.93,-3.29) .. controls (6.95,-1.4) and (3.31,-0.3) .. (0,0) .. controls (3.31,0.3) and (6.95,1.4) .. (10.93,3.29)   ;

\draw (42,94) node [anchor=north west][inner sep=0.75pt]   [align=center] {Fluid};
\draw (100.5,79) node [anchor=north west][inner sep=0.75pt]  [font=\footnotesize]  {$\delta P$};
\draw (99,103.5) node [anchor=north west][inner sep=0.75pt]  [font=\footnotesize]  {$P_{TD}$};
\draw (148,94) node [anchor=north west][inner sep=0.75pt]   [align=center] {IPUT (M)};
\draw (241,79) node [anchor=north west][inner sep=0.75pt]  [font=\footnotesize]  {$\delta u,\delta \sigma \ $};
\draw (225.5,103.5) node [anchor=north west][inner sep=0.75pt]  [font=\footnotesize]  {$ \begin{array}{l}
\delta u+\frac{\partial u}{\partial P} P_{TD} ,\ \\
\delta \sigma +\frac{\partial \sigma }{\partial P} P_{TD}
\end{array}$};
\draw (329,94) node [anchor=north west][inner sep=0.75pt]   [align=left] {IPUT (O)};
\draw (435,79) node [anchor=north west][inner sep=0.75pt]  [font=\footnotesize]  {$\delta l,\delta n\ $};
\draw (407,103.5) node [anchor=north west][inner sep=0.75pt]  [font=\footnotesize]  {$ \begin{array}{l}
\delta l+\frac{\partial l}{\partial u}\frac{\partial u}{\partial P} P_{TD} ,\\
\delta n+\frac{\partial n}{\partial \sigma }\frac{\partial \sigma }{\partial P} P_{TD}
\end{array}$};

\end{tikzpicture}   
    \caption{Transfer function and noise contribution representation in IPUT. M and O, respectively, describe the mechanical and optical components of the IPUT. $\delta P$, $\delta u$, $\delta \sigma$, $\delta l$, and $\delta n$, respectively, represent the changes in pressure, membrane displacement (axial), stresses in the membrane, waveguide's length change, and its refractive index change. $P_{TD}$ is the pressure due to the thermal displacement noise at the membrane.}
    \label{fig:IPUT_transfer}
\end{figure*}

The relation between the incoming pressure and the optical intensity---the only optical quantity we can measure using a detector---can be described via the flow diagram shown in Figure~\ref{fig:IPUT_transfer}.
Here, the IPUT is separated into purely mechanical (IPUT (M)) and purely optical (IPUT (O)) components, and their interaction is represented using their independent parameters. The incoming ultrasound pulse from the fluid exerts a pressure load ($P$) on the IPUT (M) that dynamically deforms the membrane resulting in the displacement of the membrane ($\bm{u}$) and the development of stresses ($\bm{\sigma}$) and strains ($\bm{\epsilon}$). The membrane's displacement further changes the total length of the attached waveguide ($\delta l$ in IPUT (O)), thus shifting the wavelength corresponding to the waveguide's optical resonant peak. The deformation also alters the cross-sectional area and shape of the waveguide, which influences the effective refractive index of the waveguide ($\delta n$). Additionally, the induced stress changes $n$ (effective refractive index) via the photoelastic effect~\cite{HUANG20031615}, which further modifies the optical path length. This optical resonance shift is converted to intensity variation of the interrogating laser, which is recorded using an optical detector. 

The output voltage at the detector further contains noise from several sources.
Figure~\ref{fig:IPUT_transfer} also shows the noise path, where the thermomechanical noise of the sensor, $P_{TD}$, transmits through the IPUT circuit. Noteworthy, the diagram only shows the transmission of $P_{TD}$, while other noises, such as the shot noise, are present in the optical and electric domains and are omitted from further discussion. As apparent from the figure, $P_{TD}$ is treated as a measurement quantity by the IPUT and hence is multiplied by the RTF when it reaches the detector. Thus, to characterize $P_{TD}$ and to obtain the RTF, we investigate the acoustic/elastic wave propagation through the IPUT.
\section{Acoustic/elastic wave propagation through the IPUT sensor}
\label{sec:wave_propagation_analysis}
Since the IPUT shown in Figure~\ref{fig:iput} is excited by an incoming pressure pulse from a fluid medium, its wave propagation is governed by acoustic and elastic wave equations.
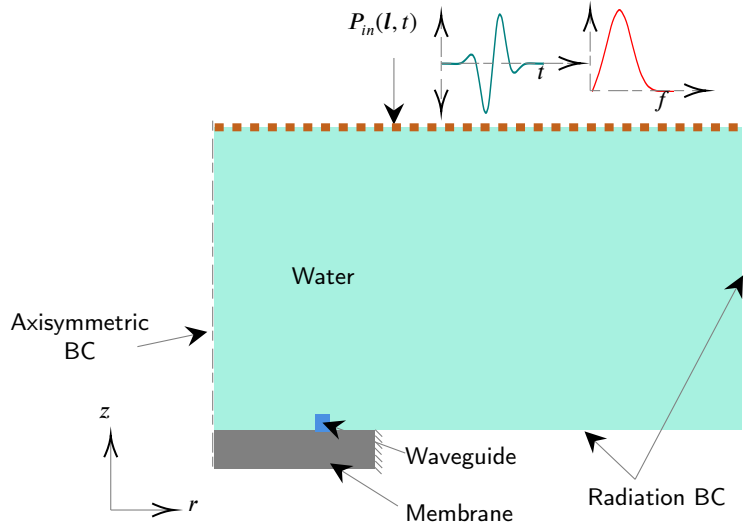
\begin{figure*}[!ht]
    \centering
    
      	\pgfplotsset{cycle list/Dark2}
	\def\svgwidth{1\linewidth}
    \centering
    \input{figures/wave_propagation_model.tex}
    \caption{Axisymmetric representation of wave propagation through IPUT when the incoming ultrasound signal is coming from water (light green region). $P(\bm{l},t)$ represents the plane Gaussian pressure pulse applied on the top edge of the water domain $\bm{l}$ marked using a brown dashed line. Radiation BCs are provided to the boundaries of the water domain to limit the reflections.}
    \label{fig:wave_propagation_analysis}
\end{figure*}
As time-domain analysis of the IPUT in 3D at high frequencies (\SI{}{\mega\hertz}) is computationally intensive when using standard computational tools such as the finite element method (FEM), we perform the analysis in 2D axisymmetric setting. Later, we also conduct a 3D time-dependent finite element analysis of the IPUT for verification.
The acoustic wave equation takes the form~\cite{pierce2019acoustics}
\begin{align}
    \Delta p={\rho_f\kappa} \ddot{p}, && \Delta \bm{v}={\rho_f\kappa} \ddot{\bm{v}},
    \label{eq:acoustic_wave_p}
\end{align}
where $p(\bm{r},t)$ and $\bm{v}(\bm{r},t)$, respectively, represent the acoustic pressure and velocity fields at the spatial coordinate $\bm{r}$ and time $t$. $\kappa$ is the compressibility of the fluid and $\rho_f$ is its mass density. $\Delta=\frac{\partial^2}{\partial r^2}+\frac{1}{r}\frac{\partial}{\partial r}$ is the Laplacian in a 2D axisymmetric system. Similarly, the elastic wave equation in the solid domain in the absence of damping and external source term takes the form~\cite{bedford2023introduction},
\begin{equation}
    \rho_s \ddot{\bm{u}}=\nabla \cdot \bm{\sigma},
    \label{eq:elastic_wave}
\end{equation}
where $\bm{u}=\{u, v, w\}$ is the displacement with $u$, $v$, $w$, respectively, representing the displacement towards $r$, $\theta$, and $z$ directions, while $\rho_s$ represents the density of the solid. $\bm{\sigma}$ is the Cauchy stress tensor in the solid material, whereas $\nabla\cdot$ is the vector divergence operator. For small perturbations (linear behavior) the stresses and strains within the solid domains are related by the following constitutive equation,
\begin{equation}
    \bm{\sigma}=\bm{C}:\bm{\gamma},
    \label{eq:constitute}
\end{equation}
 where $\bm{\gamma}$ is the strain tensor (both $\bm{\sigma}$ and $\bm{\gamma}$ are rank 2), while $\bm{C}$ is the rank 4 elasticity tensor. Since field variables are different in water and IPUT domains, we couple them using a kinematic interface condition, which ensures the continuity of normal velocity across the interface as follows,
 \begin{equation}
    \bm{n}\cdot\nabla p = -\rho_s \bm{n}\cdot\ddot{\bm{u}},
    \label{eq:acoustic_elastic_interface}
\end{equation}
where $\nabla$ is the gradient operator and $\bm{n}$ is the unit outward normal vector along the interface. Additionally, we need a dynamic interface condition to ensure the continuity of the traction,
\begin{equation}
    -p \bm{n}=\mu_s \frac{\partial \bm{u}}{\partial \bm{n}}+(\lambda_s+\mu_s)(\nabla\cdot\bm{u})\bm{n}.
    \label{eq:acoustic_elastic_interface2}
\end{equation}
The details of these interface conditions can be found in~\cite{Bao2018}. To solve this boundary value problem (BVP), it still needs necessary boundary conditions (BCs). We provide a Dirichlet BC (prescribed pressure) along the top boundary of the fluid domain as marked using a brown dashed line in Figure~\ref{fig:wave_propagation_analysis}, which takes the following form:
\begin{equation}
    P(\bm{l},t) = \bar{P} \psi(t), \hspace{3mm} \text{applied at} \hspace{3mm} \underline{r} = \underline{l},
    \label{eq:Dirichlet}
\end{equation}
where $\bar{P}$ is the constant pressure amplitude multiplied by the function $\psi(t)$ that determines the time response. The time and frequency responses of $\psi(t)$ are shown above the water domain in Figure~\ref{fig:wave_propagation_analysis}. In actual practice, the water domain is very large compared to the IPUT. To mimic this effect, we supply a plane wave radiation BC on all remaining boundaries of the fluid domain as shown in the same figure. Using Equations~\eqref{eq:acoustic_wave_p} to~\eqref{eq:Dirichlet} (wave equations, interface, and boundary conditions), we can perform a time-dependent analysis of the IPUT (M) system, which provides us with instantaneous values for the independent variables (pressure, velocity, and displacement). In the next section, we describe how these variables are used to determine the RTF.
\section{Receive transfer function calculation of the IPUT sensor}
\label{sec:transfer}
As shown in Figure~\ref{fig:IPUT_transfer}, RTF relates the incoming pressure to the wavelength/phase of the IPUT. The time-domain analysis of the IPUT (M) provides us with the displacements ($\bm{u}$) and stresses ($\bm{\sigma}$) of the waveguide. These parameters are then used to determine the IPUT's optical behavior. To that end, the following assumptions are made:
\begin{itemize}
    \item Both the membrane and waveguide are considered thin structures, i.e., the effects of rotary inertia and shear deformation are neglected;
    \item Both waveguide and membrane follow linear dynamics.
\end{itemize}
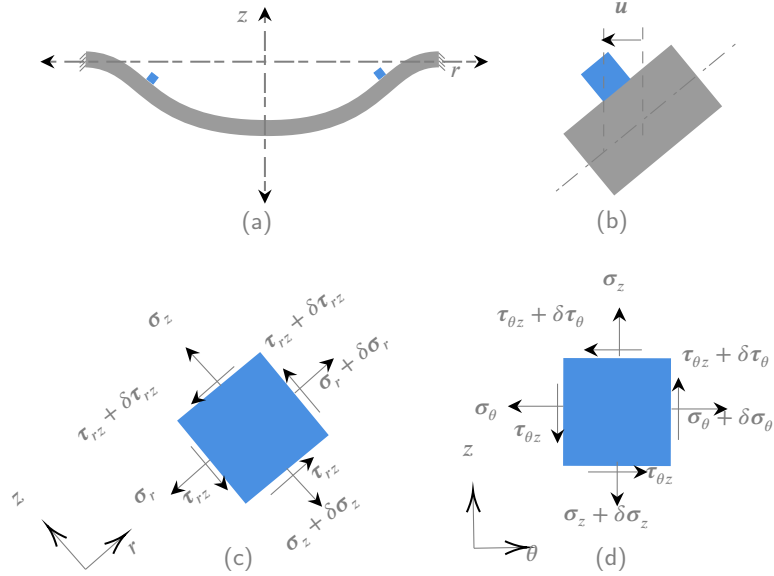
\begin{figure*}[!ht]
    \centering
      \pgfplotsset{cycle list/Dark2}
	\def\svgwidth{1\linewidth}
    \centering
    \input{figures/membrane_deformation.tex}
    \caption{Deformation of the membrane resulting in the displacement and stress of the waveguide. (a) Membrane deformed in its fundamental mode, (b) the axial displacement experienced by the waveguide due to bending of the membrane, (c) and (d), respectively, are the free body diagrams showing the normal and shear stresses in $r, z$ and $\theta, z$ planes.}
    \label{fig:membrane_deform}
\end{figure*}
Figure~\ref{fig:membrane_deform}(a) shows the behavior of the membrane-waveguide system under the influence of applied pressure. Due to the bending of the membrane, the waveguide displaces in radial and axial directions. However, since the waveguide is considered a perfect ring, to obtain its circumferential change (total length), we only need the radial displacement as marked in Figure~\ref{fig:membrane_deform}(b). In linear dynamics, the radial displacement $\bm{u}$ is directly proportional to the slope of the membrane $\frac{\partial w}{\partial r}$; thus, a higher slope yields a larger radial displacement. Figures~\ref{fig:membrane_deform}(c) and \ref{fig:membrane_deform}(d), respectively, show the stresses of the waveguide in $r,z$ and $\theta,z$ planes due to the applied pressure. We can calculate all these displacement and stress values from the time-domain analysis discussed in the previous section. The relation between the displacements and stresses on the optical behavior is discussed next.
\subsection{Wavelength change of the IPUT}
The wavelength of the light traveling through the waveguide can be related to its physical length by the following expression~\cite{reed2004silicon}:
\begin{equation}
    \lambda=\frac{L}{m}n_{\text{eff}},
    \label{eq:wavelength}
\end{equation} where $\lambda$ is the optical wavelength, $m$ is the optical mode number, $L$ is the total length of the waveguide, and $n_{\text{eff}}$ is the waveguide's effective refractive index, which is related to the optical wavefield. 
The change in the wavelength can be decomposed into changes in the refractive index and the total length as:
\begin{equation}
d\lambda = \frac{n_{\text{eff}}}{m}d L+\frac{L}{m}\left(d n_{\text{eff}}+\frac{\partial n_{\text{eff}}}{\partial\lambda}d\lambda\right),
    \label{eq:wavelength_change}
\end{equation}
where $\partial$ is the partial differential operator. The waveguide's effective index is determined by different factors such as the width $w$ and height $h$ of the waveguide, the refractive index of the material $n_{mat}$, and the optical resonant wavelength. We can further expand Eq.~\eqref{eq:wavelength_change} as follows:
\begin{equation}
    d\lambda-\frac{L}{m}\frac{\partial n_{\text{eff}}}{\partial\lambda}d\lambda=\frac{n_{\text{eff}}}{m}dL+\frac{L}{m}\left(\frac{\partial n_{\text{eff}}}{\partial w}d w+\frac{\partial n_{\text{eff}}}{\partial h}d h+\sum_i\frac{\partial n_{\text{eff}}}{\partial n_{\text{\text{mat, i}}}}d n_{\text{\text{mat, i}}}\right),
    \label{eq:wavelength_change2}   
\end{equation}      

Substituting $n_g=\left(n_{\text{eff}}-\lambda\frac{\partial n_{\text{eff}}}{\partial\lambda}\right)$, where $n_g$ is the group index in the above equation,
\begin{equation}
   d\lambda= \frac{\lambda}{n_g}\left(n_{\text{eff}} \frac{dL}{L}+\frac{\partial n_{\text{eff}}}{\partial w}d w+\frac{\partial n_{\text{eff}}}{\partial h}d h+\sum_i\frac{\partial n_{\text{eff}}}{\partial n_{\text{mat, i}}}d n_{\text{mat, i}}\right),
    \label{eq:wavelength_change5}
\end{equation}

Eq.~\eqref{eq:wavelength_change5} can be used to obtain the transfer function due to the radial displacement change, as follows:
\begin{equation}
    \left(\frac{d\lambda}{d P}\right)_{\text{el}}=\frac{\lambda n_{\text{eff}}d L}{n_g L d P},
    \label{eq:wavelength_change_length}
\end{equation}
where $d L = 2\pi u$ ($u$ is the radial component of the displacement) is the length change due to  $d P$, which is the change in the applied pressure. 
$n_{\text{eff}}$ and $n_g$ are obtained by the eigenvalue analysis of the optical waveguide. The influence of $n_{\text{eff}}$ on the wavelength is more complicated since it is influenced by the geometry and material properties as shown in Eq.~\eqref{eq:wavelength_change5}. The width and height changes are related to Poisson's effect, which can be estimated from the displacement of the waveguide using the time-domain analysis. However, the change in $n_{\text{eff}}$ due to the change in $n_{mat}$ is related to the photoelastic effect, which determines the effect of applied load on the refractive index of the material~\cite{narasimhamurty2012photoelastic}, as discussed next.
\subsection{Photoelastic effect of the IPUT}
Light propagating through an optical waveguide is governed by Maxwell's equations. With the absence of a source term, the equations take the form~\cite{huray2009maxwell}:
\begin{align}
    \nabla\times \tilde E=-\mu\frac{\partial \tilde H}{\partial t}, && \nabla\times \tilde H=\epsilon\frac{\partial \tilde E}{\partial t},
    \label{eq:light}
\end{align}
where $\tilde E$ and $\tilde H$, respectively, are the electric and magnetic fields, $\mu$ is the permeability and $\epsilon$ is the permittivity of the medium. The permittivity or dielectric tensor in 3D takes the following form~\cite{PhysRevLett.68.3603}:
\begin{equation}
    \epsilon = \begin{bmatrix}
n_{xx}^2 & n_{xy}^2 & n_{xz}^2\\
n_{xy}^2 & n_{yy}^2 & n_{yz}^2\\
n_{xz}^2 & n_{yz}^2 & n_{zz}^2\\
\end{bmatrix},
\label{eq:dielectric}
\end{equation}
where $n_{ij}$ for $i$, $j$ $\in \{x, y, z\}$ are refractive indices. Due to loading (internally or externally), stresses develop within the waveguide resulting in changes in these refractive indices, which phenomenon is known as the photoelastic effect. Photoelastic coefficients for silicon and other common materials have been calculated in literature~\cite{https://doi.org/10.1002/pssa.200671121}; however, they are still incomplete and we do not have all components of the anisotropic photoelastic tensor. The literature only reports the values for strain-optics constants $p_{11}$ and $p_{12}$. Thus the relation between the refractive index, $\bm{n}$, and strain $\bm{\gamma}$ for an isotropic point group is used, which can be expressed as~\cite{Xu1992AcoustoopticD}
\begin{equation}
    \begin{Bmatrix}
        1/n_{xx}^2\\
        1/n_{yy}^2\\
        1/n_{zz}^2\\
        1/n_{yz}^2\\
        1/n_{xz}^2\\
        1/n_{xy}^2\\
    \end{Bmatrix}
    =\begin{Bmatrix}
        1/n_0^2\\
        1/n_0^2\\
        1/n_0^2\\
        0\\
        0\\
        0\\
    \end{Bmatrix}-
    \begin{bmatrix}
        p_{11} & p_{12} & p_{12} & 0 & 0 & 0\\
        p_{12} & p_{11} & p_{12} & 0 & 0 & 0\\
        p_{12} & p_{12} & p_{11} & 0 & 0 & 0\\
        0 & 0 & 0 & p_{44} & 0 & 0\\
        0 & 0 & 0 & 0 & p_{44} & 0\\
        0 & 0 & 0 & 0 & 0 & p_{44}\\
    \end{bmatrix}
    \begin{Bmatrix}
        \gamma_{xx}\\
        \gamma_{yy}\\
        \gamma_{zz}\\
        \gamma_{yz}\\
        \gamma_{xz}\\
        \gamma_{xy}\\
    \end{Bmatrix},
    \label{eq:photo_elastic}
\end{equation}
where $p_{44}=\frac{1}{2}(p_{11}-p_{12})$ and $n_0$ is the isotropic refractive index. This relation can be expressed in terms of stresses by using the constitutive relations:
\begin{equation}
    \bm{n}_{ij} = \bm{n}_0-\bm{C}_{m}\bm{\sigma}_{ij},
 \label{eq:stress_optic}
\end{equation}
where $\bm{C_m}$ for $m\in \{1,2,3\}$ are the stress-optic constants whose values can be obtained by the following expressions,
\begin{align}
    C_1=n_0^3\left(p_{11}-2\nu p_{12}\right)/2E && C_2=n_0^3\left[p_{12}-\nu\left( p_{11}-p_{12}\right)/2E\right]/2E && C_3=n_0^3p_{44}/2G
\label{eq:stress_optics}
\end{align}
where $E$ is the Young's modulus, $\nu$ is the Poisson's ratio, and $G$ is the shear modulus.

The stress values from the time-domain acoustic analysis can be used in Eq.~\eqref{eq:stress_optic} to obtain the variations in the refractive index of the waveguide material(s). This new refractive index can then be used to solve the Maxwell equations (Eq.~\eqref{eq:light}) to derive the $n_{\text{eff}}$ of the waveguide for the supplied geometry. The variations in the $n_{\text{eff}}$ produce a wavelength shift of the waveguide as per Eq.~\eqref{eq:wavelength_change5}. The resulting transfer function can be expressed as follows:
\begin{equation}
    \left(\frac{d\lambda}{d P}\right)_{\text{ref}}=\frac{\lambda}{n_g}\frac{d n_{\text{eff}}}{d P}.
    \label{eq:wavelength_change_n_eff}
\end{equation}
Thus, the IPUT's RTF due to the incoming pressure wave is the combination of the RTF due to elongation and photoelastic effect, which can be expressed as:
\begin{equation}
    \left(\frac{d\lambda}{d P}\right)=\left(\frac{d\lambda}{d P}\right)_{\text{el}}+\left(\frac{d\lambda}{d P}\right)_{\text{ref}}= \frac{\lambda}{n_g L}\left(n_{\text{eff}}\frac{d L}{d P}+L\frac{d n_{\text{eff}}}{d P}\right).
    \label{eq:RTF}
\end{equation}

Noteworthy, wavelength change due to the waveguide's cross-sectional geometry change and due to the vacuum wavelength change have significantly less influence and are not considered in the model~\cite{6657704}.
\section{Noise equivalent pressure of the IPUT sensor}
The NEP of the IPUT (or any transducer) can be generally defined using the following expression:
\begin{equation}
     NEP = p_N/RTF,
    \label{eq:NEP_}
\end{equation}
where $p_N$ is the measured noise pressure. The NEP corresponds to the minimum detectable pressure where the pressure amplitude is $1\sigma$ ($\sigma$ standard deviation) noise level.
In the following, we derive the expression for the noise equivalent pressure generated by the thermal noise ($NEP_T$).
\label{subsec:NEP_th}
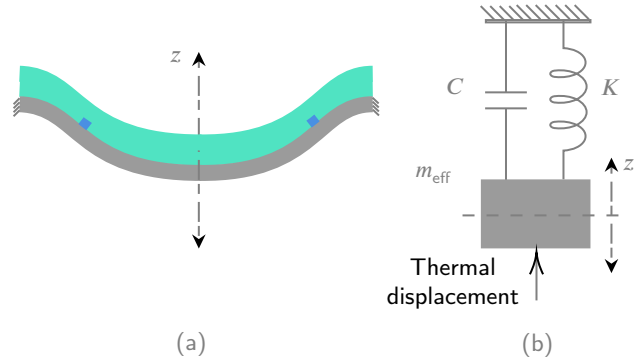
\begin{figure}[!ht]
    \centering    
    \pgfplotsset{cycle list/Dark2}
	\def\svgwidth{1\linewidth}
    \centering
    \input{figures/noise_model}
    \caption{IPUT with the surrounding fluid represented as a single harmonic oscillator (a) Membrane deformation (with the fluid on top) dominated by its fundamental mode, (b) Motion of the IPUT in the presence of surrounding fluid medium represented as a weakly damped single harmonic oscillator with $K$ stiffness, $C$ damping factor, and $m_{\text{eff}}$ the effective mass of the oscillator. The thermal displacement from the fluid acts as the source term for the membrane.}
    \label{fig:noise_model}
\end{figure}
Thermal noise is generated by losses in the sensor and Brownian motion in the surrounding medium~\cite{macdonald2006noise}. The noise contribution in the output response due to thermal noise is derived using the equipartition theorem~\cite{H_-J_Butt_1995}, while also incorporating complex geometric features and multiple materials. 
The following assumptions are used in addition to the ones discussed in Section~\ref{sec:transfer}:
\begin{itemize}
    \item The excitation of the membrane is only via the thermal displacement of the membrane, i.e., no other source term is present in the system;
    \item The motion of the membrane is assumed to follow the behavior of a weakly damped single harmonic oscillator (see Figure~\ref{fig:noise_model}(b));
\end{itemize}
The following considerations are also made to incorporate the influence of the surrounding fluid medium on the IPUT:
\begin{itemize}
    \item The inertia effect of the fluid load is incorporated in the effective mass of the resonator;
    \item The resonator's damping factor includes the dissipation due to the acoustic radiation to the fluid;
    \item The effects of fluid on the resonance frequency of the resonator is also consolidated in the resonator's stiffness.
\end{itemize}
According to the equipartition theorem, the energy is shared equally amongst all energetically accessible degrees of freedom (DOFs), and each quadratic DOF will, on average, possess an energy of $\frac{1}{2}k_BT$, where $k_B$ is the Boltzmann's constant whose value in SI unit is \SI[per-mode=symbol]{1.38e-23}{\joule\per\kelvin} and $T$ is the absolute temperature in Kelvin. In the case of the micro-mechanical resonator, we use a displacement DOF, $\bm{U}$ to represent its motion, which in a linear regime can be expressed as the product of spatial and temporal functions.
\begin{equation}
    \bm{U}(\bm{x},t) = \phi(t)\bar{\bm{u}}(\bm{x}),
    \label{eq:disp_decompose}
\end{equation}
where, $\bar{\bm{u}}$ is the mode shape corresponding to the desired motion of the membrane, while $\phi$ is the time-dependent function describing the membrane's motion. The mode shape can be computed by the eigenvalue analysis of the membrane, which depending on the complexity of the geometry, can be performed analytically or numerically.
$\phi(t)$ can be obtained by solving the differential equation of motion (DEOM) of the membrane. For simplicity, we represent the IPUT membrane as a single harmonic oscillator with stiffness $K$, damping factor $C$, and the effective mass $m_{\text{eff}}$ as shown in Figure~\ref{fig:noise_model}. The DEOM then can be represented as follows:
\begin{equation}
    m_{\text{eff}}\ddot{\phi}+C\dot{\phi}+K \phi =F(t).
    \label{eq:DEOM}
\end{equation}
Here $m_{\text{eff}}$ incorporates the influence of the fluid and can be expressed as follows:
\begin{equation}
    m_{\text{eff}}= 0.3\times (1+\beta)m_s
    \label{eq:eff_mass}
\end{equation}
where $m_s$ is the mass of the sensor and $\beta=0.65381\frac{\rho_f\times r_s}{\rho_s\times t_s}$ is the fluid mass contribution factor~\cite{AMABILI1996743}.
$F(t)$ is the applied force, which can be expressed as the product of the applied pressure $p(t)$ and the cross-sectional area $A_s$. The $NEP_{T}$ for the resonator is the ratio of the resonator's spectral response due to the thermal Langevin force divided by the pressure sensitivity of the resonator~\cite{POMEAU2017570}. The Langevin force represents the fluctuating part of the Brownian motion, which acts as the source term for the resonator's motion. Since the generated thermal noise is a Gaussian white noise, it can be separated into different frequency components (spectral response).
The sensitivity of the resonator due to the applied pressure can be obtained by solving Eq.~\eqref{eq:DEOM} in the frequency domain and dividing it by the input pressure (also in the frequency domain):
\begin{equation}
    \Sigma(\omega)=\Phi(\omega)/P(\omega),
    \label{eq:sensitivity}
\end{equation}
where $\Sigma(\omega)$ is the sensitivity expressed in \SI[per-mode=symbol]{}{\meter\per\pascal}. Here $\Phi(\omega)$ and $P(\omega)$, respectively, are Fourier transforms of the displacement, $\phi(t)$, and pressure $p(t)$. To obtain the resonator's response due to thermal force by the equipartition theorem, we need to express the mean-square amplitude of motion in terms of the power spectral density (PSD). The PSD is obtained by squaring the frequency spectrum of the resonator's motion ($\Phi(\omega)$) and dividing it by the resolution bandwidth (RBW---the frequency range of the resonator's response bounded by \SI{-6}{\decibel} signal amplitude). To obtain the PSD, an autocorrelation function $R(t)$ is used, which relates the signal amplitude to itself at a later time~\cite{HAUER2013181}.
\begin{equation}
    R(t) = \lim_{T_0 \to \infty} \frac{1}{T_0}\int_0^{T_0} \phi(t')\phi(t'+t)dt',
    \label{eq:autocorrelation}
\end{equation}
where $T_0$ is the time window. The two-sided PSD, $\tilde{P}(\omega)$, is the Fourier transform of the autocorrelation function~\cite{norton2003fundamentals}. Similarly, $R(t)$ is the inverse Fourier transform of $\tilde{P}(\omega)$.
\begin{align}
    \tilde{P}(\omega) = \int_{-\infty}^{\infty}e^{i\omega t}R(t) dt && R(t) = \frac{1}{2\pi}\int_{-\infty}^{\infty}e^{-i\omega t}\tilde{P}(\omega) d\omega.
    \label{eq:two_sided_PSD}  
\end{align}
In real situations, $\tilde{P}(\omega)$ is an even function and thus can be expressed using the positive frequency range such that the single-sided PSD, $S(\omega)=2 \tilde{P}(\omega)$~\cite{press2007numerical}. The mean-square amplitude of the motion is defined as:
\begin{equation}
    \langle \phi^2(t)\rangle = \frac{1}{T_0}\int_0^{T_0} [\phi(t)]^2dt,
    \label{eq:mean_square}
\end{equation}
which is identical to $R(t)$ at zero time shift provided the signal is sampled for a sufficiently long time window. Using Eq.\eqref{eq:two_sided_PSD}, $\langle a^2(t)\rangle$ can be expressed in terms of the PSD~\cite{doi:10.1063/1.347347}.
\begin{equation}
    \langle \phi^2(t)\rangle =  \frac{1}{2\pi}\int_0^{\infty}S(\omega) d\omega.
    \label{eq:mean_square_PSD}
\end{equation}
By taking the Fourier transform of Eq.~\eqref{eq:DEOM}, we can express $S(\omega)$ in terms of $S_F(\omega)$, which is the PSD due to the forcing function $F(t)$:
\begin{equation}
    S(\omega)=|\chi|^2 S_F(\omega),
    \label{eq:forcing_PSD}
\end{equation}
where $\chi$ is the mechanical susceptibility obtained as:
\begin{equation}
    \chi(\omega) = \frac{\mathcal{F}(\phi(t))}{\mathcal{F}(F(t))}=\frac{1}{m_{\text{eff}}(\omega_0^2-\omega^2+i\omega\omega_0/Q)^{0.5}},
    \label{eq:susceptibility}
\end{equation}
where $\omega_0$ and $Q$ are the natural frequency and the Q-factor of the resonator (corresponding to the required mode). Since $F(t)$ is generated by the thermal noise, which is broadband, $S_F(\omega)$ can be considered to be a constant thermal force $S_F^{T}$. By using \Crefrange{eq:mean_square_PSD}{eq:susceptibility}, we can write:
\begin{equation}
    \langle \phi^2(t)\rangle = \frac{S_F^{T}}{2\pi m_{\text{eff}}^2}\int_0^{\infty}\frac{d\omega}{(\omega^2-\omega_0^2)^2+(\omega\omega_0/Q)^2} = \frac{S_F^{T} Q}{4\omega_0^3 m_{\text{eff}}^2}.
    \label{eq:amplitude_thermal_force}
\end{equation}
Using the equipartition theorem, we represent the time-averaged potential energy $\tilde{U}$ of the resonator as follows:
\begin{equation}
    \langle \tilde{U}\rangle =\frac{1}{2}m_{\text{eff}}\omega_0^2\langle \phi^2(t)\rangle =\frac{1}{2}k_B T.
    \label{eq:equipartition}
\end{equation}
From Eqs.~\eqref{eq:amplitude_thermal_force} and \eqref{eq:equipartition}, we get the expression of $S_F^{T}$ as:
\begin{equation}
    S_F^{T}=\frac{4k_BT\omega_0 m_{\text{eff}}}{Q},
    \label{eq:thermal_force}
\end{equation}
and the resonator's PSD as:
\begin{equation}
    S(\omega)=\frac{4k_BT\omega_0}{m_{\text{eff}}Q[(\omega-\omega_0)^2+(\omega\omega_0/Q)^2]}.
    \label{eq:resonator_thermal_force}
\end{equation}

As mentioned previously, to get the NEP, we also need to obtain the pressure sensitivity ($\Sigma(\omega)$) of the resonator. $\Sigma(\omega)$ can be obtained by using the mechanical susceptibility (refer Eq.~\eqref{eq:susceptibility}) as follows:
\begin{equation}
    \Sigma(\omega)=\frac{A_s}{m_{\text{eff}}[(\omega-\omega_0)^2+(\omega\omega_0/Q)^2]^{0.5}},
    \label{eq:sensitivity1}
\end{equation}
where $A_s$ is the cross-sectional area of the sensor. Now, from Eqs.~\eqref{eq:resonator_thermal_force} and \eqref{eq:sensitivity1}, we obtain thermal-noise-induced pressure of the sensor as follows:
\begin{equation}
    p_s = \frac{\sqrt{S}}{\Sigma}= \sqrt{\frac{4k_BT\omega_0 m_{\text{eff}}}{QA_s^2}}
    \label{eq:NEP_T}
\end{equation}
As shown in Figure~\ref{fig:IPUT_transfer} and Section~\ref{sec:definition}, $p_s$ gets multiplied by the RTF (Eq.~\eqref{eq:RTF}) at the receiver. Hence, the equivalent of $p_s$ in wavelength ($NEP_\lambda$) or phase ($NEP_\phi$) is used to represent the thermal noise as follows:
\begin{equation}
    NEP_{\lambda s} =\left(\frac{d\lambda}{dP}\right)p_s = \frac{\lambda}{n_g L}\left(n_{\text{eff}}\frac{d L}{d P}+L\frac{d n_{\text{eff}}}{d P}\right) \sqrt{\frac{4k_BT\omega_0 m_{\text{eff}}}{QA_s^2}}.
    \label{eq:NElambda_T}
\end{equation}    
To minimize $p_s$, we need to maximize $Q$ and $A_s$ while minimizing $m_{\text{eff}}$, and $\omega_0$ of the membrane. In the case of medical ultrasound imaging, the penetration depth decides the required central frequency, and the axial resolution of the image dictates the frequency bandwidth; hence, these parameters cannot be modified to improve $p_s$. The remaining parameters $m_{\text{eff}}$ and $A_s$ are also not independent; for instance, increasing $A_s$ (without changing the thickness) would increase $m_{\text{eff}}$, but also decreases $\omega_0$ and the bandwidth. \hl{To study the sensor's performance under various parameters}, we investigate the influence of variations in geometry and material properties on $p_s$. These effects are incorporated via $m_{\text{eff}}$. 
\subsection{Incorporating complex geometric and material behavior via the effective mass}
The effective mass of each resonator mode can be obtained from its potential energy. For the $n^{th}$ mode, each volume element $dV$ will have a potential energy corresponding to the harmonic oscillator with a mass element $dm=\rho(\bm{x})dV$ given by:
\begin{equation}
    d\tilde{U} =\frac{1}{2}\rho(\bm{x})\omega_n^2|\phi_n(t)\bm{u}_n(\bm{x})|^2dV
    \label{eq:potential_elemental}
\end{equation}
The total potential energy of the mode is given by integrating over the entire volume of the device, $V$.
\begin{equation}
    \Tilde{U}=\frac{1}{2}\omega_n^2|\phi_n(t)|^2\int\rho(\bm{x})|\bm{u}_n(\bm{x})|^2 dV = \frac{1}{2}\omega_n^2|\phi_n(t)|^2 m_{eff,} 
    \label{eq:potential_complete}
\end{equation}
where $m_{\text{eff}}$ is defined as:
\begin{equation}
    m_{\text{eff}}=\int \rho(\bm{x})|\bm{u}_n(\bm{x})|^2 dV.
    \label{eq:effective_mass}
\end{equation}
Thus, with the normalized mode shape $\bm{u}_n(\bm{x})$, $m_{\text{eff}}$ can be calculated. We can numerically obtain the normalized mode for a complex geometry (e.g., via FEM), which can be used to determine its effective mass. Additionally, the density, $\rho$ is also a function of space; thus, multiple layers can also be included in the membrane aiding us in exploring different possibilities to minimize $NEP_T$.
Note that the expression of NEP (Equations~\eqref{eq:NEP_T}) is in terms of \SI[per-mode=symbol]{}{\pascal\per\sqrt{\hertz}}, which needed to be multiplied by the square root of the operation bandwidth to obtain the NEP in \SI{}{\pascal}. 
The expressions of RTF and NEP complete the IPUT model, and we proceed to the model validation.
\section{Validation of RTF and NEP models}
\begin{figure*}[!ht]
   	\pgfplotsset{cycle list/Dark2}
	\def\svgwidth{0.9\linewidth}
    \centering
    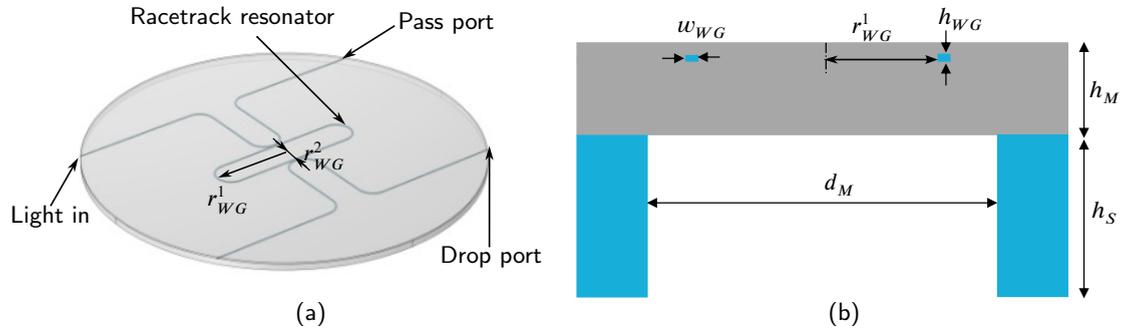
    \caption{The IPUT design from \cite{westerveld2014silicon} where (a) is the isometric view and (b) shows the side view of the IPUT only considering the racetrack waveguide (excluding all ports) with the substrate backing. Here the waveguide width $w_{WG}=\SI{400}{\nano\meter}$, its height $h_{WG}=\SI{220}{\nano\meter}$, the diameter of the membrane $d_{M}=\SI{124}{\micro\meter}$, its height $h_{M}=\SI{2.5}{\micro\meter}$, and the height of the substrate $h_{S}=\SI{250}{\micro\meter}$. The major radius of the racetrack is $r_{WG}^1=\SI{18}{\micro\meter}$ while its minor radius is $r_{WG}^2=\SI{5}{\micro\meter}$. The input (laser coming in), pass, and drop ports are also marked in (a).}
    \label{fig:Wouter_device}
\end{figure*}

To validate the RTF and NEP models, we compare various properties of the IPUT namely, the resonance frequency, Q-factor, RTF, and NEP with the literature. To that end, we select the design discussed in \cite{Leinders2015, westerveld2014silicon}, shown in Figure~\ref{fig:Wouter_device}.
The membrane is composed of SiO\textsubscript{2} and has a diameter of \SI{124}{\micro\meter} and a thickness of \SI{2}{\micro\meter}. A silicon racetrack waveguide of \SI{400}{\nano\meter} width and \SI{220}{\nano\meter} height is attached to the membrane, which has the major radius of $r_{WG}^1=\SI{18}{\micro\meter}$ and minor radius of $r_{WG}^2=\SI{5}{\micro\meter}$. A SiO\textsubscript{2} cladding of \SI{0.5}{\micro\meter} is then provided on top to protect the waveguide from the environment. Thus the total thickness of the membrane is \SI{2.5}{\micro\meter}.


After performing the wave propagation analysis discussed in Section~\ref{sec:wave_propagation_analysis} we obtain its dynamic response. To that end, a time-domain analysis for 40 cycles is performed with a time step of T/15. T is the period corresponding to $f_{max}=1.5\times f_c$, where $f_c$ is the central frequency of the input pulse.
\subsection{Time- and frequency-domain behavior of the IPUT}
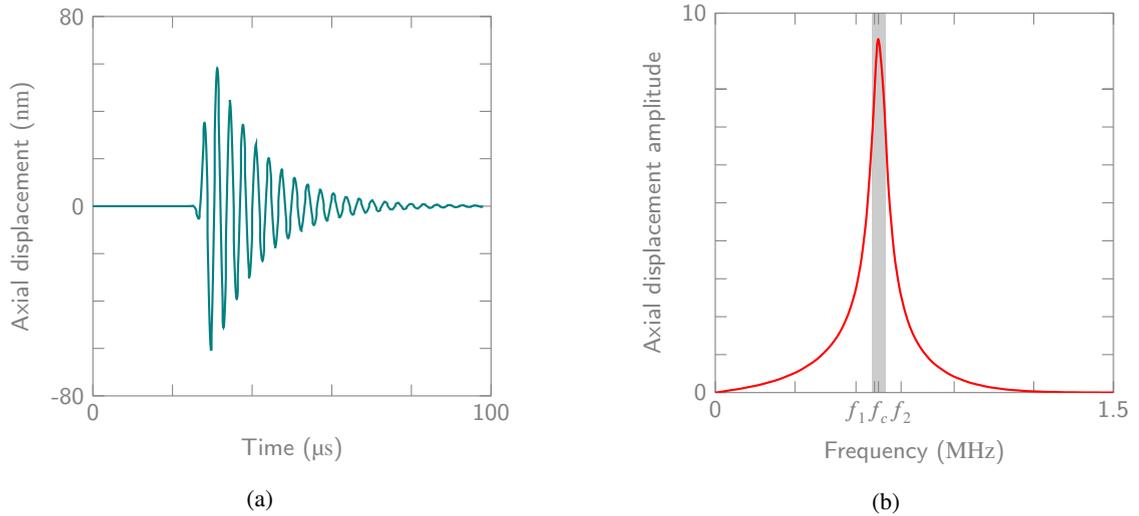
\begin{figure}[!ht]      
          \centering
\begin{minipage}{0.495\textwidth}
	\centering
	\pgfplotsset{cycle list/Dark2}
	\def\svgwidth{1\linewidth}
  \input{figures/disp_time.tex}
	\subcaption{}
	\label{fig:membrane_time}
\end{minipage}
\begin{minipage}{0.495\textwidth}
	\centering
	\pgfplotsset{cycle list/Dark2}
	\def\svgwidth{1\linewidth}
  \input{figures/disp_freq.tex}
	\subcaption{}
	\label{fig:membrane_freq}
\end{minipage}
\caption{The output displacement signal is represented as functions of (a) time and (b) frequency. The frequencies that correspond to the peak displacement ($f_c=\SI{0.615}{\mega\hertz}$), and the lower ($f_1=\SI{0.59}{\mega\hertz}$) and upper ($f_2=\SI{0.642}{\mega\hertz}$) bounds of \SI{-3}{\decibel} BW are also marked in (b).}
\label{fig:disp_out}
\end{figure}
The axial displacement of the membrane extracted at its center is plotted in Figure~\ref{fig:disp_out}(a), which behaves similarly to a damped harmonic oscillator. The frequency response is obtained by transforming the time response to the frequency domain via Fourier transform (FFT) as shown in Figure~\ref{fig:disp_out}(b). Here the \SI{-3}{\decibel} bandwidth (BW) also called full-width half maximum (FWHM) is shown using a gray-shaded box bounded by $f_1=\SI{0.59}{\mega\hertz}$ and $f_2=\SI{0.642}{\mega\hertz}$. The frequency corresponding to the peak displacement provides its resonant frequency, which is $f_c=\SI{0.615}{\mega\hertz}$. The experimentally obtained resonance frequency from \cite{Leinders2015} is \SI{0.76}{\mega\hertz}, which is about \SI{19}{\percent} higher than the numerical predictions. This difference is due to the geometric variations during the fabrication (diameter and thickness change due to over/under etching of the membrane) and the effect of prestress as discussed in \cite{8724528}. 
We calculate the Q-factor from this frequency response using the Lorentzian as follows:
\begin{equation}
    Q=\frac{f_c}{f_2-f_1}=11.75,
    \label{eq:Q_factor}
\end{equation}
resulting in a \SI{-3}{\decibel} BW of \SI{8.5}{\percent}. The Q-factor obtained experimentally is around 10~\cite{westerveld2014silicon}, which is \SI{15}{\percent} lower than the model prediction. We further estimate the output transfer function of the IPUT by using the displacement of the waveguide as discussed next.
\subsection{Receive transfer function of the IPUT}
Since the exerted pressure deforms the waveguide along with the membrane, we calculate the RTF of the IPUT due to the wavelength change and the photoelastic effect as discussed in Section~\ref{sec:transfer}. To obtain the wavelength change we need to determine the waveguide's physical length change, which is influenced by its radial displacement. 
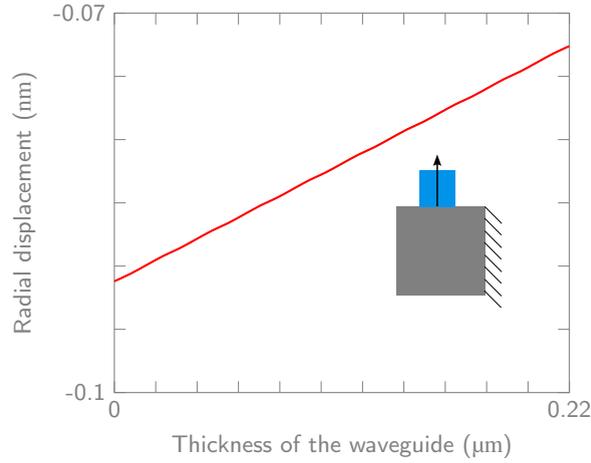
\begin{figure}[!ht]      
          \centering
	\centering
	\pgfplotsset{cycle list/Dark2}
	\def\svgwidth{0.75\linewidth}
  \input{figures/radial_disp_}
	\caption{Radial displacement extracted along the thickness of the waveguide. The inset shows the schematic of the waveguide and membrane region where the arrow goes through the center of the waveguide and represents the abscissa.}
	\label{fig:waveguide_radial}
\end{figure}
The radial displacement as a function of time (similar to Figure~\ref{fig:membrane_time}) is extracted along the thickness of the waveguide (see the inset of Figure~\ref{fig:waveguide_radial}). These time responses are then transformed into the frequency domain (like Figure~\ref{fig:membrane_freq}) from which the displacements corresponding to the resonant frequency $f_c$ are selected and plotted against the distance from the bottom of the waveguide (through its height) as shown in Figure~\ref{fig:waveguide_radial}. Using this radial displacement, we obtain the RTF. To that end, $n_{\text{eff}}=2.228$, $n_g=4.386$ and $\lambda=\SI{1550}{\nano\meter}$ are selected, while $\Delta P$ is the incoming pressure value corresponding to the resonant frequency (taken from the input pulse). $n_{\text{eff}}$ and $n_g$ are calculated via eigenvalue analysis of the waveguide using FIMMWAVE (an optical mode solver), while $\Delta P$ is obtained from the time-domain acoustic analysis. By using Eq.~\eqref{eq:wavelength_change_length}, we calculate the RTF due to the wavelength shift as $\left(\frac{\Delta \lambda}{\Delta P}\right)_{\text{el}} = \SI[per-mode=symbol]{-27}{\femto\meter\per\pascal}$.

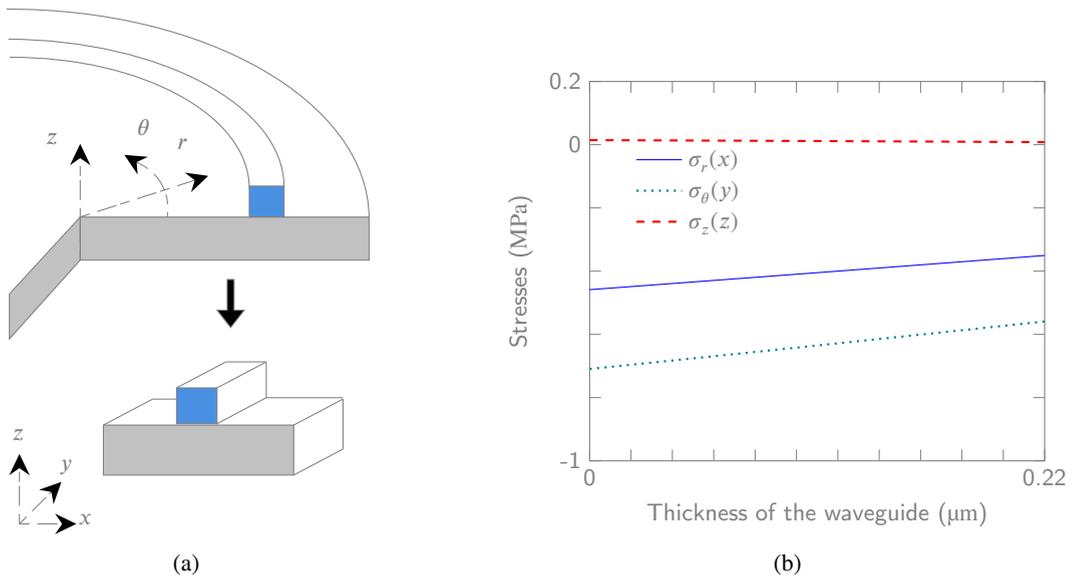
\begin{figure}
   \centering
   \begin{subfigure}{0.475\textwidth}
	\centering
	\pgfplotsset{cycle list/Dark2}
	\def\svgwidth{1\linewidth}
  \input{figures/coordinate_transform.tex}
	\subcaption{}
	\label{fig:coordinate_transformation}
\end{subfigure}
\begin{subfigure}{0.475\textwidth}
    \centering
	\pgfplotsset{cycle list/Dark2}
	\def\svgwidth{1\linewidth}
  \input{figures/stresses.tex}
  \subcaption{}
	\label{fig:stress_waveguide}
\end{subfigure}
\caption{Normal stresses on the waveguide measured along the vertical edge of the waveguide. (a) coordinate transformation from cylindrical to Cartesian coordinate system and (b) corresponding stresses extracted along the waveguide thickness, where the axial stress is acting in the opposite direction and has a considerably lower magnitude than radial and angular values.}
\end{figure}

RTF of the IPUT due to stresses is more complicated since the stresses are rank 2 tensors as opposed to displacements (vectors). Additionally, the waveguide is defined in the cylindrical coordinate system (refer Figures~\ref{fig:membrane_deform}(c) and~\ref{fig:membrane_deform}(d)) while the photoelastic tensor is in the Cartesian coordinate system (see Eq.~\eqref{eq:photo_elastic}), so we need to transform the waveguide to Cartesian coordinates. Since the stresses are evaluated locally, we can assume that for small angles, the waveguide is straight as shown in Figure~\ref{fig:coordinate_transformation}. 
The stresses towards $r$ (radial), $\theta$ (angular), and $z$ (axial) directions can be used, respectively, as stresses in $x$, $z$, and $y$ directions. 
The resulting relation between different normal stresses and the waveguide thickness is provided in Figure~\ref{fig:stress_waveguide}. The values of shear stresses were negligible compared to the normal stresses and hence are neglected. Additionally, the axial stress ($\bm{\sigma_z}$) is approximately two orders of magnitude lower than the angular and radial stresses. Hence, the effect of Poisson's ratio on the wavelength shift would also be lower. Using the stress values and photoelastic tensor from Eq.~\eqref{eq:photo_elastic}, we can calculate the transfer function due to the stresses using Eq.~\eqref{eq:wavelength_change_n_eff}. Here \hl{$C_1=\SI{-1.25e-11}{\per\pascal}$ and $C_2=\SI{4.66e-12}{\per\pascal}$} are selected for silicon~\cite{https://doi.org/10.1002/pssa.200671121} and resulting wavelength shift due to the stress is $\left(\frac{\Delta\lambda}{\Delta P}\right)_{\text{ref}}=\SI[per-mode=symbol]{-6.53}{\femto\meter\per\pascal}$. Thus the total RTF of the IPUT due to the elongation and photoelasticity is \SI[per-mode=symbol]{-33.6}{\femto\meter\per\pascal}. The total wavelength shift obtained experimentally in \cite{westerveld2014silicon} is \SI[per-mode=symbol]{67}{\femto\meter\per\pascal}, which is twice the model prediction. This could be due to the effects of geometric variation during fabrication (for instance change in membrane thickness by over-etching), and the prestress.
\subsection{Comparison of the IPUT's RTF with a 3D model}
Since the IPUT geometry provided in Figure~\ref{fig:Wouter_device}(a) is not of an axisymmetric device, we also check its response using a 3D acousto-mechanical model.
\begin{figure}[!ht]
 \centering
      \pgfplotsset{cycle list/Dark2}
	\def\svgwidth{0.4\linewidth}
    \centering
    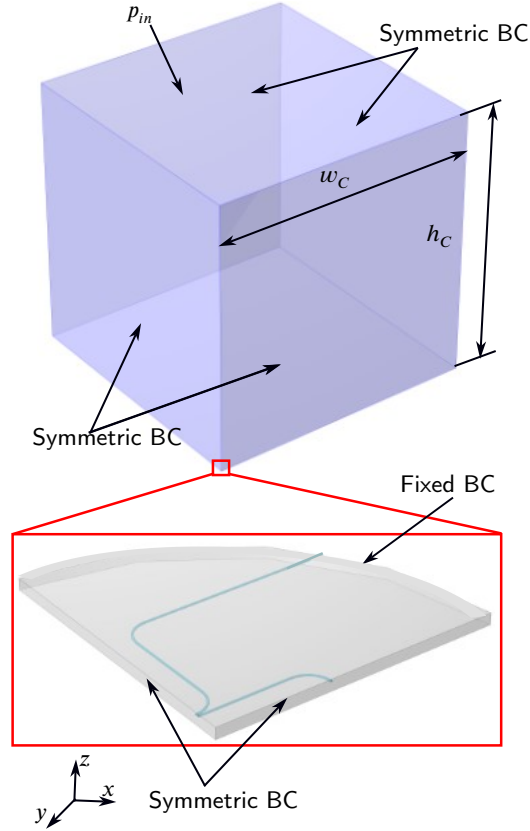
    \caption{Schematic representation of the 3D IPUT model, where the blue region represents the fluid domain while the gray and teal (embedded in gray) regions, respectively, are the membrane and waveguide. Since the RR is quarter-symmetric, symmetry BCs are provided on the two flat surfaces of the membrane and waveguide system. Additionally, for the fluid domain, symmetric BCs are supplied on all flat surfaces perpendicular to the $x-y$ plane, while the top surface is provided with a prescribed pressure ($P_{in}$) load, and the bottom surface is left free. The width of the fluid domain is $w_C=\SI{6.25}{\milli\meter}$ while the height $h_C=\SI{25}{\milli\meter}$.}
    \label{fig:3D_model}
\end{figure}
The 3D analysis is very similar to the 2D axisymmetric modeling discussed in Section~\ref{sec:wave_propagation_analysis}, except that waves propagating in all directions are considered and the BCs are modified. Here, instead of an axisymmetric BC, symmetric BCs are provided to the solid domain (two flat surfaces) and fluid domain (four flat surfaces normal to $x-y$ plane) as shown in Figure~\ref{fig:3D_model}. The time-domain analysis is performed for 40 cycles with a time step of T/15 (similar to the axisymmetric model).
\begin{figure}[!ht]      
          \centering
\begin{minipage}{0.495\textwidth}
	\centering
	\pgfplotsset{cycle list/Dark2}
	\def\svgwidth{1\linewidth}
  \input{figures/disp_time.tex}
	\subcaption{}
	\label{fig:membrane_time_3D}
\end{minipage}
\begin{minipage}{0.495\textwidth}
	\centering
	\pgfplotsset{cycle list/Dark2}
	\def\svgwidth{1\linewidth}
  \input{figures/disp_freq.tex}
	\subcaption{}
	\label{fig:membrane_freq_3D}
\end{minipage}
\caption{The output displacement signal computed for the 3D model is represented as functions of (a) time and (b) frequency. The frequencies correspond to the peak displacement ($f_c^*=\SI{0.623}{\mega\hertz}$), the lower ($f_1^*=\SI{0.594}{\mega\hertz}$) and upper ($f_2^*=\SI{0.648}{\mega\hertz}$) bounds of the \SI{-3}{\decibel} BW are also marked in (b). See Figure~\ref{fig:membrane_freq} for corresponding values in 2D axisymmetric case.}
\label{fig:disp_out_3D}
\end{figure}
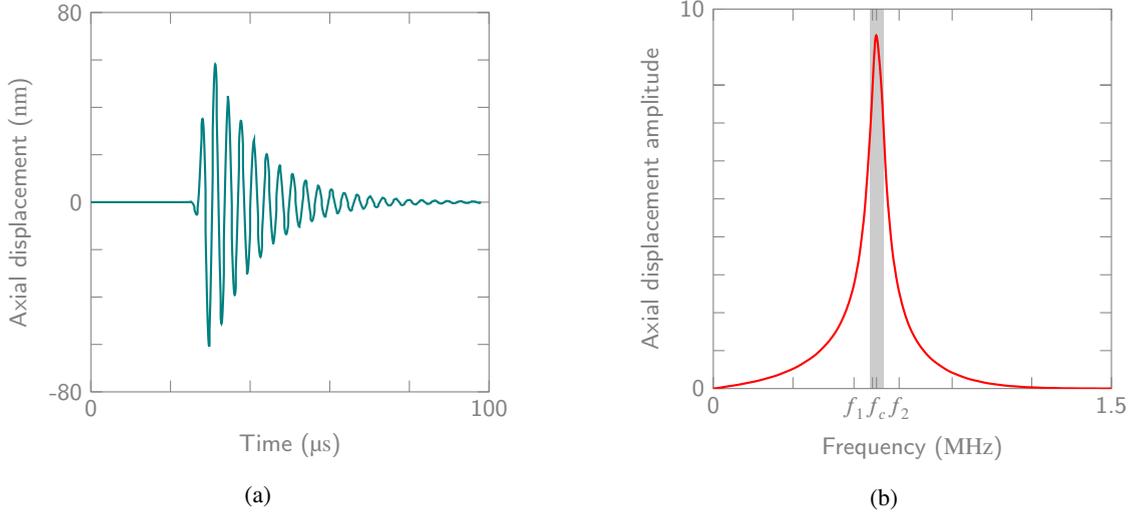
The resulting time and frequency response of the axial displacement are provided in Figures~\ref{fig:membrane_time_3D}, and \ref{fig:membrane_freq_3D}, respectively.
Since the dynamic response of the IPUT is dominated by the membrane, its resonance frequency and Q factor are comparable with the 2D axisymmetric analysis and we observe that the resonance frequency is \SI{0.623}{\mega\hertz} (\SI{1.3}{\percent} higher than corresponding 2D axisymmetric case) while the Q-factor is 11.5 (\SI{2.2}{\percent} lower than 2D axisymmetric case). 
\begin{figure}[!ht]      
          \centering
	\centering
\begin{minipage}{0.495\textwidth}
	\centering
	\pgfplotsset{cycle list/Dark2}
	\def\svgwidth{1\linewidth}
  \input{figures/radial_disp_3D}
	\subcaption{}
	\label{fig:radial_disp_3D}
\end{minipage}
\begin{minipage}{0.495\textwidth}
	\centering
	\pgfplotsset{cycle list/Dark2}
	\def\svgwidth{1\linewidth}
  \input{figures/stresses_3D}
	\subcaption{}
	\label{fig:stress_3D}
\end{minipage}
	\caption{(a) Radial displacement extracted along the waveguide. The inset shows the schematic of the quarter waveguide where the arrow represents the direction along which the displacement is extracted. (b) Different stresses of the waveguide extracted along its thickness (the inset shows the schematic of the region of the waveguide and membrane).}
	\label{fig:waveguide_disp_stress_3D}
\end{figure}
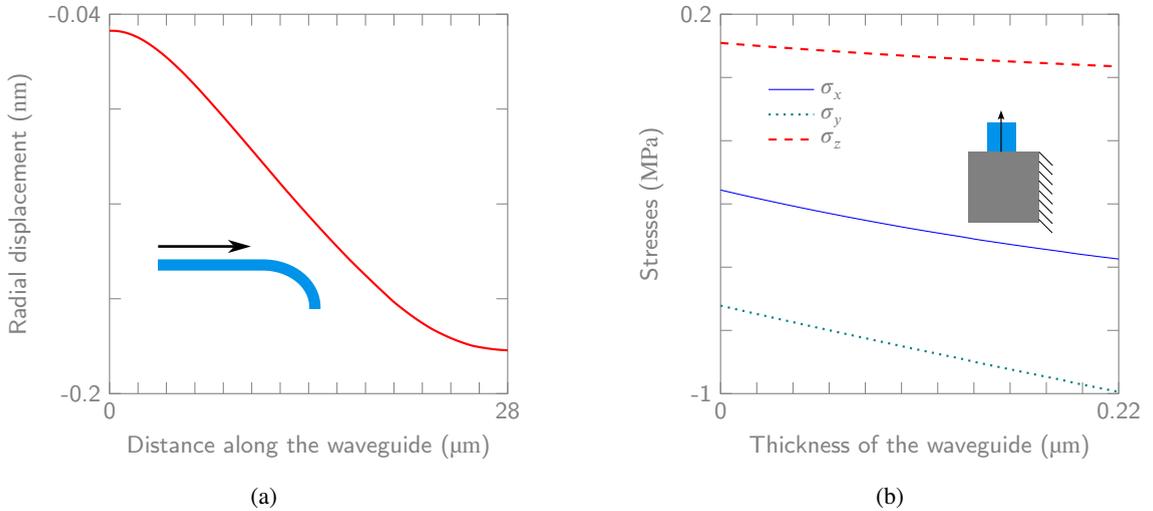
The RTF is calculated by using the radial displacement of the waveguide shown in Figure~\ref{fig:radial_disp_3D} for which the quarter racetrack is used (see the inset of Figure~\ref{fig:radial_disp_3D}). The displacement is then multiplied by four to get the total radial expansion of the ring. Similarly, the stresses are also calculated along the waveguide's height as shown in Figure~\ref{fig:stress_3D}, which is used to obtain the variations in $n_{\text{eff}}$. The RTF obtained from the 3D model by combining both elongation $\left(\frac{\Delta \lambda}{\Delta P}\right)_{\text{el}}=\SI[per-mode=symbol]{-19}{\femto\meter\per\pascal}$ and photoelastic effects $\left(\frac{\Delta \lambda}{\Delta P}\right)_{\text{ref}} =\SI[per-mode=symbol]{-3}{\femto\meter\per\pascal}$ is \SI[per-mode=symbol]{-22}{\femto\meter\per\pascal} at \SI{0.623}{\mega\hertz}. Although the resonance frequency and Q-factor are comparable between the 3D and 2D axisymmetric models, the RTF obtained here is about \SI{35}{\percent} lower than the 2D axisymmetric case. This is because, in the former, the actual racetrack geometry of the optical ring characterized by $R_{WG}^1$ and $R_{WG}^2$ from Figure~\ref{fig:Wouter_device}(a) is considered, while in the latter, a circular ring with the radius same as $R_{WG}^1$ of the racetrack is used (see Figure~\ref{fig:Wouter_device}(b)). This leads to a larger length change in the waveguide and thereby higher RTF in the case of the 2D axisymmetric analysis. Nevertheless, both 2D axisymmetric and 3D analysis predicts lower RTF values (factor of 2 for 2D and 3 for 3D) than that of the experiments, which could be due to the influence of prestress on the IPUT.
\subsection{Effects of prestress on the response of the IPUT}
Since the prestress has a significant influence on the resonance frequency of the IPUT~\cite{7926285}, we expect a similar influence on the RTF as well. Additionally, as discussed in~\cite{8724528}, the device from Figure~\ref{fig:Wouter_device}(a) had experienced buckling due to the prestress. This could result in a change in the mode shape of the IPUT, hence significantly affecting the radial displacement of the waveguides leading to a change in the RTF, which is beyond the scope of this paper. Now we proceed to compare the NEP with the model predictions.
\subsection{NEP comparison}
\begin{figure}[!ht]      
          \centering
\begin{minipage}{0.495\textwidth}
	\centering
	\pgfplotsset{cycle list/Dark2}
	\def\svgwidth{1\linewidth}
  \input{figures/NEP_freq.tex}
	\subcaption{}
	\label{fig:NEP_freq}
\end{minipage}
\begin{minipage}{0.495\textwidth}
	\centering
	\pgfplotsset{cycle list/Dark2}
	\def\svgwidth{1\linewidth}
  \input{figures/NEP_Q.tex}
	\subcaption{}
	\label{fig:NEP_Q}
\end{minipage}
\caption{Plots of $NEP_{th}$ as a function of (a) frequency and (b) Q factor for the operational ranges of medical ultrasound applications. The red triangles in both plots indicate the experimentally obtained $NEP$ from \cite{Leinders2015}. It can be seen that the IPUT is still limited by the noise from the photodetector for the medical ultrasound applications (frequency range from \SI{100}{\kilo\hertz} to \SI{5}{\mega\hertz} and Q factor from 1 to 15.)}
\end{figure}
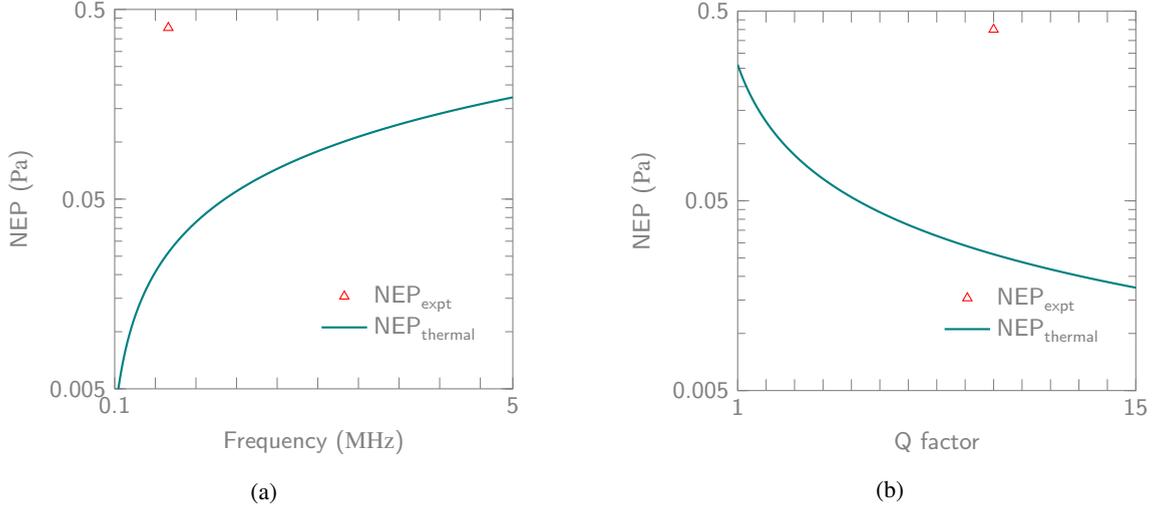
The NEP of the IPUT is calculated using Eqs.~\eqref{eq:NEP_T}, where all parameters are already obtained. The IPUT's NEP is \SI{0.0787}{\milli\pascal\per\sqrt{\hertz}}. For a FWHM BW (\SI{148}{\kilo\hertz}), $\text{NEP}=\SI{0.03}{\pascal}$. The total NEP value shown in \cite{Leinders2015} is \SI{0.4}{\pascal}, which is more than 1 order higher than the thermal noise of the IPUT. This is because, in the selected work, the thermal noise has not been explicitly extracted experimentally and the noise floor of the system is limited by the noise of the read-out system's amplifier. Figures~\ref{fig:NEP_freq} and \ref{fig:NEP_Q}, respectively, show the variation of NEP with frequency and Q factor. Figure~\ref{fig:NEP_freq} assumes a constant Q factor of 10 (experimental value) and the NEP increases with the resonance frequency of the mechanical resonator. Similarly, Figure~\ref{fig:NEP_Q} assumes a constant resonance frequency of \SI{0.76}{\mega\hertz} (also experimental value), and here the NEP decreases with the increase in Q factor. Nevertheless, in both cases, the experimentally obtained NEP is much higher than the thermal noise for the entire operational frequency and Q factor ranges. Thus, to further improve the performance of the IPUT, a low-noise read-out system is needed. 
\section{Conclusions}
In this work, novel semi-analytical models were proposed to characterize the receive transfer function (RTF) and noise equivalent pressure (NEP) of integrated photonic ultrasound transducers (IPUTs) operating underwater. The models were validated by comparing them with the literature. We draw the following conclusions.
\begin{itemize}
    \item To accurately characterize the RTF of an IPUT, it is necessary to incorporate its dynamic behavior into the acoustic model;
    \item To design IPUTs for the best performance, it is necessary to characterize their noise behavior along with RTF since the ratio between these parameters (signal-to-noise ratio) determines the IPUT's NEP. For applications such as medical imaging where the devices are limited by the signal-to-noise-ratio (the transmitted pulse's amplitude has an upper limit due to safety reasons), a low NEP is preferable for their better performance;
    \item To accurately characterize the thermal noise of the IPUT, a precise description of the RTF (Eq.~\eqref{eq:RTF}) is also needed as the thermal noise interacts with the outside world (detector) via the IPUT's RTF;
    \item Among the different factors contributing to the RTF, the waveguide's elongation and photoelastic effect have major influences;
    \item In the present case, the two predominant effects that influence the IPUT's RTF namely, the optical waveguide's physical elongation and the photoelastic effect act in the same direction due to the state of the stress tensor. In the case of a uniaxial loading of the optical waveguide, the aforementioned effects act in opposite directions as reported in the literature~\cite{6657704};
    \item The thermal noise from the fluid medium can be neglected compared to the same from the IPUT sensor since the magnitude of the former is much lower than the latter; 
\end{itemize}
The model presented in this study can be extended to include complex geometric/material properties to extend the IPUT's design space further. Additionally, we can determine various parameters using this model, which can be tuned to optimize the IPUT's performance. A future direction is to use this model to design a novel IPUT device with high RTF and low NEP and validate its performance experimentally. Another interesting direction is to investigate the influence of prestress on the resonance frequency, Q-factor, RTF, and NEP of the IPUT.
\section*{Acknowledgement}
The authors greatly appreciate the support from the funding partner Nederlandse Organisatie voor Wetenschappelijk Onderzoek (NWO) for the grant entitled ``Opto acoustic sensor and ultrasonic microbubbles for dosimetry in proton therapy'' with the grant number NWA-1160.18.095.
\printcredits
\section*{Data availability}
The raw data supporting this study's findings are available from the corresponding author on request.
\bibliography{cas-refs}

\end{document}

%% file: figures/IPUT_geometry.pdf_tex
\begingroup%
  \makeatletter%
  \providecommand\color[2][]{%
    \errmessage{(Inkscape) Color is used for the text in Inkscape, but the package 'color.sty' is not loaded}%
    \renewcommand\color[2][]{}%
  }%
  \providecommand\transparent[1]{%
    \errmessage{(Inkscape) Transparency is used (non-zero) for the text in Inkscape, but the package 'transparent.sty' is not loaded}%
    \renewcommand\transparent[1]{}%
  }%
  \providecommand\rotatebox[2]{#2}%
  \newcommand*\fsize{\dimexpr\f@size pt\relax}%
  \newcommand*\lineheight[1]{\fontsize{\fsize}{#1\fsize}\selectfont}%
  \ifx\svgwidth\undefined%
    \setlength{\unitlength}{285.65853882bp}%
    \ifx\svgscale\undefined%
      \relax%
    \else%
      \setlength{\unitlength}{\unitlength * \real{\svgscale}}%
    \fi%
  \else%
    \setlength{\unitlength}{\svgwidth}%
  \fi%
  \global\let\svgwidth\undefined%
  \global\let\svgscale\undefined%
  \makeatother%
  \begin{picture}(1,0.32547689)%
    \lineheight{1}%
    \setlength\tabcolsep{0pt}%
    \put(0.04223154,0.28041112){\makebox(0,0)[lt]{\lineheight{5}\smash{\begin{tabular}[t]{l}Mechanical membrane\end{tabular}}}}%
    \put(0.6330855,0.27783303){\makebox(0,0)[lt]{\lineheight{5}\smash{\begin{tabular}[t]{l}Optical waveguide\end{tabular}}}}%
    \put(0,0){\includegraphics[width=\unitlength,page=1]{IPUT_geometry.pdf}}%
  \end{picture}%
\endgroup%

%% file: figures/wave_propagation_model.tex
\begin{tikzpicture}[x=0.75pt,y=0.75pt,yscale=-1,xscale=1]

\draw  [draw opacity=0][fill={rgb, 255:red, 80; green, 227; blue, 194 }  ,fill opacity=0.5 ] (194.67,78.33) -- (458.33,78.33) -- (458.33,229.52) -- (194.67,229.52) -- cycle ;
\draw  [draw opacity=0][fill={rgb, 255:red, 128; green, 128; blue, 128 }  ,fill opacity=1 ] (194.67,229.52) -- (275,229.52) -- (275,249) -- (194.67,249) -- cycle ;
\draw  [draw opacity=0][fill={rgb, 255:red, 74; green, 144; blue, 226 }  ,fill opacity=1 ] (245,221.67) -- (252.09,221.67) -- (252.09,230.33) -- (245,230.33) -- cycle ;
\draw    (275,229.52) -- (278.48,233) ;
\draw    (275,232.52) -- (278.48,236) ;
\draw    (274.67,238.52) -- (278.14,242) ;
\draw    (275,235.52) -- (278.48,239) ;
\draw    (275,242.19) -- (278.48,245.67) ;
\draw    (274.67,248.19) -- (278.14,251.67) ;
\draw    (275,245.19) -- (278.48,248.67) ;
\draw  [dash pattern={on 3.75pt off 3pt on 7.5pt off 1.5pt}]  (193.67,75) -- (193.67,249) ;
\draw    (284.33,43.67) -- (284.33,74) ;
\draw [shift={(284.33,77)}, rotate = 270] [fill={rgb, 255:red, 0; green, 0; blue, 0 }  ][line width=0.08]  [draw opacity=0] (10.72,-5.15) -- (0,0) -- (10.72,5.15) -- (7.12,0) -- cycle    ;
\draw    (286.33,267.67) -- (254.87,247.93) ;
\draw [shift={(252.33,246.33)}, rotate = 32.11] [fill={rgb, 255:red, 0; green, 0; blue, 0 }  ][line width=0.08]  [draw opacity=0] (10.72,-5.15) -- (0,0) -- (10.72,5.15) -- (7.12,0) -- cycle    ;
\draw    (290,241.19) -- (251.36,227.03) ;
\draw [shift={(248.55,226)}, rotate = 20.12] [fill={rgb, 255:red, 0; green, 0; blue, 0 }  ][line width=0.08]  [draw opacity=0] (10.72,-5.15) -- (0,0) -- (10.72,5.15) -- (7.12,0) -- cycle    ;
\draw    (405,254.33) -- (381.84,232.4) ;
\draw [shift={(379.67,230.33)}, rotate = 43.45] [fill={rgb, 255:red, 0; green, 0; blue, 0 }  ][line width=0.08]  [draw opacity=0] (10.72,-5.15) -- (0,0) -- (10.72,5.15) -- (7.12,0) -- cycle    ;
\draw    (405,254.33) -- (457.6,154.98) ;
\draw [shift={(459,152.33)}, rotate = 117.9] [fill={rgb, 255:red, 0; green, 0; blue, 0 }  ][line width=0.08]  [draw opacity=0] (10.72,-5.15) -- (0,0) -- (10.72,5.15) -- (7.12,0) -- cycle    ;
\draw (333.83,46.5) node  {\includegraphics[width=37.75pt,height=37.75pt]{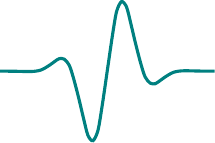}};
\draw  [dash pattern={on 3.75pt off 3pt on 7.5pt off 1.5pt}]  (308,46.33) -- (375.67,46.33) ;
\draw [shift={(377.67,46.33)}, rotate = 180] [color={rgb, 255:red, 0; green, 0; blue, 0 }  ][line width=0.75]    (10.93,-3.29) .. controls (6.95,-1.4) and (3.31,-0.3) .. (0,0) .. controls (3.31,0.3) and (6.95,1.4) .. (10.93,3.29)   ;
\draw  [dash pattern={on 3.75pt off 3pt on 7.5pt off 1.5pt}]  (308,24.17) -- (308,68.5) ;
\draw [shift={(308,70.5)}, rotate = 270] [color={rgb, 255:red, 0; green, 0; blue, 0 }  ][line width=0.75]    (10.93,-3.29) .. controls (6.95,-1.4) and (3.31,-0.3) .. (0,0) .. controls (3.31,0.3) and (6.95,1.4) .. (10.93,3.29)   ;
\draw [shift={(308,22.17)}, rotate = 90] [color={rgb, 255:red, 0; green, 0; blue, 0 }  ][line width=0.75]    (10.93,-3.29) .. controls (6.95,-1.4) and (3.31,-0.3) .. (0,0) .. controls (3.31,0.3) and (6.95,1.4) .. (10.93,3.29)   ;

\draw    (142.67,269.67) -- (142.67,234.33) ;
\draw [shift={(142.67,232.33)}, rotate = 90] [color={rgb, 255:red, 0; green, 0; blue, 0 }  ][line width=0.75]    (10.93,-3.29) .. controls (6.95,-1.4) and (3.31,-0.3) .. (0,0) .. controls (3.31,0.3) and (6.95,1.4) .. (10.93,3.29)   ;
\draw    (142.67,269.67) -- (172.33,269.67) ;
\draw [shift={(174.33,269.67)}, rotate = 180] [color={rgb, 255:red, 0; green, 0; blue, 0 }  ][line width=0.75]    (10.93,-3.29) .. controls (6.95,-1.4) and (3.31,-0.3) .. (0,0) .. controls (3.31,0.3) and (6.95,1.4) .. (10.93,3.29)   ;
\draw [color={rgb, 255:red, 194; green, 98; blue, 28 }  ,draw opacity=1 ][line width=3]  [dash pattern={on 3.38pt off 3.27pt}]  (194.67,78.33) -- (458.33,78.33) ;
\draw (404,40) node  {\includegraphics[width=30.75pt,height=30.75pt]{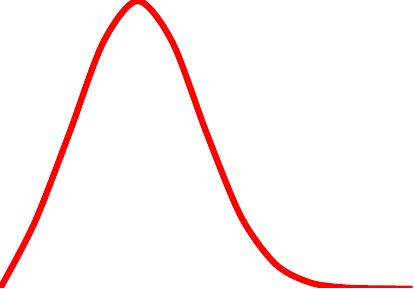}};
\draw  [dash pattern={on 3.75pt off 3pt on 7.5pt off 1.5pt}]  (382.5,60) -- (438.5,60) ;
\draw [shift={(440.5,60)}, rotate = 180] [color={rgb, 255:red, 0; green, 0; blue, 0 }  ][line width=0.75]    (10.93,-3.29) .. controls (6.95,-1.4) and (3.31,-0.3) .. (0,0) .. controls (3.31,0.3) and (6.95,1.4) .. (10.93,3.29)   ;
\draw  [dash pattern={on 3.75pt off 3pt on 7.5pt off 1.5pt}]  (382.5,21.5) -- (382.5,60) ;
\draw [shift={(382.5,19.5)}, rotate = 90] [color={rgb, 255:red, 0; green, 0; blue, 0 }  ][line width=0.75]    (10.93,-3.29) .. controls (6.95,-1.4) and (3.31,-0.3) .. (0,0) .. controls (3.31,0.3) and (6.95,1.4) .. (10.93,3.29)   ;

\draw    (156,188) -- (188.06,181.22) ;
\draw [shift={(191,180.6)}, rotate = 168.06] [fill={rgb, 255:red, 0; green, 0; blue, 0 }  ][line width=0.08]  [draw opacity=0] (10.72,-5.15) -- (0,0) -- (10.72,5.15) -- (7.12,0) -- cycle    ;

\draw (259.67,19.67) node [anchor=north west][inner sep=0.75pt][ align=left]   [color={rgb, 255:red, 0; green, 0; blue, 0 }  ,opacity=1] {$\displaystyle P_{in}(\boldsymbol{l} ,t)$};
\draw (232,147) node [anchor=north west][inner sep=0.75pt]   [align=left][color={rgb, 255:red, 0; green, 0; blue, 0 }  ,opacity=1] {Water};
\draw (288.67,265) node [anchor=north west][inner sep=0.75pt] [color={rgb, 255:red, 0; green, 0; blue, 0 }  ,opacity=1]  [align=left] {Membrane};
\draw (288.67,236.67) node [anchor=north west][inner sep=0.75pt] [color={rgb, 255:red, 0; green, 0; blue, 0 }  ,opacity=1]  [align=left] {Waveguide};
\draw (372.67,257.33) node [anchor=north west][inner sep=0.75pt] [color={rgb, 255:red, 0; green, 0; blue, 0 }  ,opacity=1]  [align=left] {\begin{minipage}[lt]{62.82pt}\setlength\topsep{0pt}
\begin{center}
Radiation BC
\end{center}

\end{minipage}};
\draw (354.83,46) node [anchor=north west][inner sep=0.75pt] [color={rgb, 255:red, 0; green, 0; blue, 0 }  ,opacity=1]  [align=left] {$\displaystyle t$};
\draw (180.33,263.67) node [anchor=north west][inner sep=0.75pt] [color={rgb, 255:red, 0; green, 0; blue, 0 }  ,opacity=1]   {$r$};
\draw (135.33,217) node [anchor=north west][inner sep=0.75pt] [color={rgb, 255:red, 0; green, 0; blue, 0 }  ,opacity=1]   {$z$};
\draw (413.33,59) node [anchor=north west][inner sep=0.75pt] [color={rgb, 255:red, 0; green, 0; blue, 0 }  ,opacity=1]  [align=left] {$\displaystyle f$};
\draw (83.67,170) node [anchor=north west][inner sep=0.75pt] [color={rgb, 255:red, 0; green, 0; blue, 0 }  ,opacity=1]  [align=left] {\begin{minipage}[lt]{63.36pt}\setlength\topsep{0pt}
\begin{center}
Axisymmetric\\BC
\end{center}

\end{minipage}};

\end{tikzpicture}

%% file: figures/membrane_deformation.tex
\begin{tikzpicture}[x=0.75pt,y=0.75pt,yscale=-1,xscale=1]

\draw [color={rgb, 255:red, 155; green, 155; blue, 155 }  ,draw opacity=1 ][line width=6]    (35.33,33.5) .. controls (66,34.5) and (63,68) .. (126.33,67.83) ;
\draw [color={rgb, 255:red, 155; green, 155; blue, 155 }  ,draw opacity=1 ][line width=6]    (125.33,67.83) .. controls (184,68.5) and (184,33.5) .. (212,33.5) ;
\draw    (35.33,30.5) -- (32.33,33.5) ;
\draw    (35.33,32.5) -- (32.33,35.5) ;
\draw    (35.33,34.5) -- (32.33,37.5) ;
\draw    (35.33,36.5) -- (32.33,39.5) ;
\draw    (211.87,30.21) -- (214.96,33.13) ;
\draw    (211.93,32.21) -- (215.01,35.12) ;
\draw    (211.99,34.21) -- (215.07,37.12) ;
\draw    (212.04,36.21) -- (215.13,39.12) ;
\draw  [color={rgb, 255:red, 74; green, 144; blue, 226 }  ,draw opacity=1 ][fill={rgb, 255:red, 74; green, 144; blue, 226 }  ,fill opacity=1 ] (179.61,40.29) -- (182.77,37.82) -- (185.24,40.98) -- (182.08,43.45) -- cycle ;
\draw  [color={rgb, 255:red, 74; green, 144; blue, 226 }  ,draw opacity=1 ][fill={rgb, 255:red, 74; green, 144; blue, 226 }  ,fill opacity=1 ] (71.57,42.62) -- (68.41,40.15) -- (65.94,43.31) -- (69.1,45.78) -- cycle ;
\draw [line width=0.75]  [dash pattern={on 3.75pt off 3pt on 7.5pt off 1.5pt}]  (15.67,34.67) -- (233.67,34.67) ;
\draw [shift={(236.67,34.67)}, rotate = 180] [fill={rgb, 255:red, 0; green, 0; blue, 0 }  ][line width=0.08]  [draw opacity=0] (7.14,-3.43) -- (0,0) -- (7.14,3.43) -- (4.74,0) -- cycle    ;
\draw [shift={(12.67,34.67)}, rotate = 0] [fill={rgb, 255:red, 0; green, 0; blue, 0 }  ][line width=0.08]  [draw opacity=0] (7.14,-3.43) -- (0,0) -- (7.14,3.43) -- (4.74,0) -- cycle    ;
\draw [line width=0.75]  [dash pattern={on 3.75pt off 3pt on 7.5pt off 1.5pt}]  (125,10.33) -- (125,102.67) ;
\draw [shift={(125,105.67)}, rotate = 270] [fill={rgb, 255:red, 0; green, 0; blue, 0 }  ][line width=0.08]  [draw opacity=0] (7.14,-3.43) -- (0,0) -- (7.14,3.43) -- (4.74,0) -- cycle    ;
\draw [shift={(125,7.33)}, rotate = 90] [fill={rgb, 255:red, 0; green, 0; blue, 0 }  ][line width=0.08]  [draw opacity=0] (7.14,-3.43) -- (0,0) -- (7.14,3.43) -- (4.74,0) -- cycle    ;
\draw  [draw opacity=0][fill={rgb, 255:red, 155; green, 155; blue, 155 }  ,fill opacity=1 ] (274.29,70.71) -- (328.28,26.15) -- (353.74,56.99) -- (299.76,101.56) -- cycle ;
\draw  [draw opacity=0][fill={rgb, 255:red, 74; green, 144; blue, 226 }  ,fill opacity=1 ] (283.05,41) -- (296.68,29.75) -- (307.92,43.38) -- (294.3,54.63) -- cycle ;
\draw  [dash pattern={on 1.5pt off 3.75pt on 7.5pt off 1.5pt}]  (360.35,25.85) -- (267.68,101.85) ;
\draw  [dash pattern={on 4.5pt off 4.5pt}]  (314.02,17.5) -- (314.02,71.5) ;
\draw  [dash pattern={on 4.5pt off 4.5pt}]  (294.49,18) -- (294.49,79.5) ;
\draw    (313.17,23.83) -- (296.17,23.83) ;
\draw [shift={(293.17,23.83)}, rotate = 360] [fill={rgb, 255:red, 0; green, 0; blue, 0 }  ][line width=0.08]  [draw opacity=0] (7.14,-3.43) -- (0,0) -- (7.14,3.43) -- (4.74,0) -- cycle    ;
\draw  [draw opacity=0][fill={rgb, 255:red, 74; green, 144; blue, 226 }  ,fill opacity=1 ] (81.08,214.13) -- (122.63,179.83) -- (156.92,221.38) -- (115.38,255.68) -- cycle ;
\draw    (97.46,233.8) -- (80.12,249.18) ;
\draw [shift={(77.88,251.18)}, rotate = 318.42] [fill={rgb, 255:red, 0; green, 0; blue, 0 }  ][line width=0.08]  [draw opacity=0] (7.14,-3.43) -- (0,0) -- (7.14,3.43) -- (4.74,0) -- cycle    ;
\draw    (157.29,184.92) -- (139.96,200.3) ;
\draw [shift={(159.54,182.93)}, rotate = 138.42] [fill={rgb, 255:red, 0; green, 0; blue, 0 }  ][line width=0.08]  [draw opacity=0] (7.14,-3.43) -- (0,0) -- (7.14,3.43) -- (4.74,0) -- cycle    ;
\draw    (88,227) -- (104.57,246.7) ;
\draw [shift={(106.5,249)}, rotate = 229.94] [fill={rgb, 255:red, 0; green, 0; blue, 0 }  ][line width=0.08]  [draw opacity=0] (7.14,-3.43) -- (0,0) -- (7.14,3.43) -- (4.74,0) -- cycle    ;
\draw    (135.93,190.3) -- (152.5,210) ;
\draw [shift={(134,188)}, rotate = 49.94] [fill={rgb, 255:red, 0; green, 0; blue, 0 }  ][line width=0.08]  [draw opacity=0] (7.14,-3.43) -- (0,0) -- (7.14,3.43) -- (4.74,0) -- cycle    ;
\draw    (103.04,196.64) -- (87.31,179.62) ;
\draw [shift={(85.28,177.41)}, rotate = 47.28] [fill={rgb, 255:red, 0; green, 0; blue, 0 }  ][line width=0.08]  [draw opacity=0] (7.14,-3.43) -- (0,0) -- (7.14,3.43) -- (4.74,0) -- cycle    ;
\draw    (151.6,254.99) -- (135.88,237.97) ;
\draw [shift={(153.63,257.2)}, rotate = 227.28] [fill={rgb, 255:red, 0; green, 0; blue, 0 }  ][line width=0.08]  [draw opacity=0] (7.14,-3.43) -- (0,0) -- (7.14,3.43) -- (4.74,0) -- cycle    ;
\draw    (109.65,187.05) -- (90.28,204.01) ;
\draw [shift={(88.02,205.99)}, rotate = 318.8] [fill={rgb, 255:red, 0; green, 0; blue, 0 }  ][line width=0.08]  [draw opacity=0] (7.14,-3.43) -- (0,0) -- (7.14,3.43) -- (4.74,0) -- cycle    ;
\draw    (147.3,234.24) -- (127.93,251.2) ;
\draw [shift={(149.56,232.27)}, rotate = 138.8] [fill={rgb, 255:red, 0; green, 0; blue, 0 }  ][line width=0.08]  [draw opacity=0] (7.14,-3.43) -- (0,0) -- (7.14,3.43) -- (4.74,0) -- cycle    ;

\draw  [draw opacity=0][fill={rgb, 255:red, 74; green, 144; blue, 226 }  ,fill opacity=1 ] (274.29,182.83) -- (328.16,182.96) -- (328.03,236.84) -- (274.16,236.71) -- cycle ;
\draw    (274.33,206.93) -- (249.5,206.93) ;
\draw [shift={(246.5,206.93)}, rotate = 360] [fill={rgb, 255:red, 0; green, 0; blue, 0 }  ][line width=0.08]  [draw opacity=0] (7.14,-3.43) -- (0,0) -- (7.14,3.43) -- (4.74,0) -- cycle    ;
\draw    (353.5,208.29) -- (328.43,208.29) ;
\draw [shift={(356.5,208.29)}, rotate = 180] [fill={rgb, 255:red, 0; green, 0; blue, 0 }  ][line width=0.08]  [draw opacity=0] (7.14,-3.43) -- (0,0) -- (7.14,3.43) -- (4.74,0) -- cycle    ;
\draw    (271.39,195.66) -- (271.39,222.5) ;
\draw [shift={(271.39,225.5)}, rotate = 270] [fill={rgb, 255:red, 0; green, 0; blue, 0 }  ][line width=0.08]  [draw opacity=0] (7.14,-3.43) -- (0,0) -- (7.14,3.43) -- (4.74,0) -- cycle    ;
\draw    (331.89,195.5) -- (331.89,223.76) ;
\draw [shift={(331.89,192.5)}, rotate = 90] [fill={rgb, 255:red, 0; green, 0; blue, 0 }  ][line width=0.08]  [draw opacity=0] (7.14,-3.43) -- (0,0) -- (7.14,3.43) -- (4.74,0) -- cycle    ;
\draw    (302.35,181.89) -- (302.35,160.5) ;
\draw [shift={(302.35,157.5)}, rotate = 90] [fill={rgb, 255:red, 0; green, 0; blue, 0 }  ][line width=0.08]  [draw opacity=0] (7.14,-3.43) -- (0,0) -- (7.14,3.43) -- (4.74,0) -- cycle    ;
\draw    (301.24,254) -- (301.24,234.67) ;
\draw [shift={(301.24,257)}, rotate = 270] [fill={rgb, 255:red, 0; green, 0; blue, 0 }  ][line width=0.08]  [draw opacity=0] (7.14,-3.43) -- (0,0) -- (7.14,3.43) -- (4.74,0) -- cycle    ;
\draw    (313.57,178.74) -- (287.5,178.74) ;
\draw [shift={(284.5,178.74)}, rotate = 360] [fill={rgb, 255:red, 0; green, 0; blue, 0 }  ][line width=0.08]  [draw opacity=0] (7.14,-3.43) -- (0,0) -- (7.14,3.43) -- (4.74,0) -- cycle    ;
\draw    (313.5,239.78) -- (286.67,239.78) ;
\draw [shift={(316.5,239.78)}, rotate = 180] [fill={rgb, 255:red, 0; green, 0; blue, 0 }  ][line width=0.08]  [draw opacity=0] (7.14,-3.43) -- (0,0) -- (7.14,3.43) -- (4.74,0) -- cycle    ;

\draw    (35.27,288.49) -- (17.46,268.2) ;
\draw [shift={(16.14,266.7)}, rotate = 48.72] [color={rgb, 255:red, 0; green, 0; blue, 0 }  ][line width=0.75]    (10.93,-3.29) .. controls (6.95,-1.4) and (3.31,-0.3) .. (0,0) .. controls (3.31,0.3) and (6.95,1.4) .. (10.93,3.29)   ;
\draw    (35.27,288.49) -- (53.68,272.33) ;
\draw [shift={(55.19,271.01)}, rotate = 138.72] [color={rgb, 255:red, 0; green, 0; blue, 0 }  ][line width=0.75]    (10.93,-3.29) .. controls (6.95,-1.4) and (3.31,-0.3) .. (0,0) .. controls (3.31,0.3) and (6.95,1.4) .. (10.93,3.29)   ;
\draw    (229.59,277.93) -- (229.14,250.93) ;
\draw [shift={(229.11,248.93)}, rotate = 89.06] [color={rgb, 255:red, 0; green, 0; blue, 0 }  ][line width=0.75]    (10.93,-3.29) .. controls (6.95,-1.4) and (3.31,-0.3) .. (0,0) .. controls (3.31,0.3) and (6.95,1.4) .. (10.93,3.29)   ;
\draw    (229.59,277.93) -- (254.08,277.52) ;
\draw [shift={(256.08,277.49)}, rotate = 179.06] [color={rgb, 255:red, 0; green, 0; blue, 0 }  ][line width=0.75]    (10.93,-3.29) .. controls (6.95,-1.4) and (3.31,-0.3) .. (0,0) .. controls (3.31,0.3) and (6.95,1.4) .. (10.93,3.29)   ;

\draw (216.67,35.67) node [anchor=north west][inner sep=0.75pt]    {$r$};
\draw (109.33,8.33) node [anchor=north west][inner sep=0.75pt]    {$z$};
\draw (112,108) node [anchor=north west][inner sep=0.75pt]   [align=left] {(a)};
\draw (289.5,107.5) node [anchor=north west][inner sep=0.75pt]   [align=left] {(b)};
\draw (103,276) node [anchor=north west][inner sep=0.75pt]   [align=left] {(c)};
\draw (289,276) node [anchor=north west][inner sep=0.75pt]   [align=left] {(d)};
\draw (56.45,251.76) node [anchor=north west][inner sep=0.75pt]  [rotate=-322.19]  {$\boldsymbol{\sigma} _{r}$};
\draw (147.45,193) node [anchor=north west][inner sep=0.75pt]  [rotate=-322.19]  {$\boldsymbol{\sigma} _{r} +\delta \boldsymbol{\sigma} _{r}$};
\draw (63.45,164) node [anchor=north west][inner sep=0.75pt]  [rotate=-322.19]  {$\boldsymbol{\sigma} _{z}$};
\draw (130.95,273.24) node [anchor=north west][inner sep=0.75pt]  [rotate=-322.19]  {$\boldsymbol{\sigma} _{z} +\delta \boldsymbol{\sigma} _{z}$};
\draw (80.95,252.5) node [anchor=north west][inner sep=0.75pt]  [rotate=-322.19]  {$\boldsymbol{\tau} _{rz}$};
\draw (145.45,240) node [anchor=north west][inner sep=0.75pt]  [rotate=-322.19]  {$\boldsymbol{\tau} _{rz}$};
\draw (120.45,172.5) node [anchor=north west][inner sep=0.75pt]  [rotate=-322.19]  {$\boldsymbol{\tau} _{rz} +\delta \boldsymbol{\tau} _{rz}$};
\draw (26.95,219) node [anchor=north west][inner sep=0.75pt]  [rotate=-322.19]  {$\boldsymbol{\tau} _{rz} +\delta \boldsymbol{\tau} _{rz}$};
\draw (292.74,139.52) node [anchor=north west][inner sep=0.75pt]  [rotate=-1.14]  {$\boldsymbol{\sigma} _{z}$};
\draw (228.74,204.02) node [anchor=north west][inner sep=0.75pt]  [rotate=-1.14]  {$\boldsymbol{\sigma} _{\theta }$};
\draw (335.24,206.02) node [anchor=north west][inner sep=0.75pt]  [rotate=-1.14]  {$\boldsymbol{\sigma} _{\theta } +\delta \boldsymbol{\sigma} _{\theta }$};
\draw (273.24,254.52) node [anchor=north west][inner sep=0.75pt]  [rotate=-1.14]  {$\boldsymbol{\sigma} _{z} +\delta \boldsymbol{\sigma} _{z}$};
\draw (248.01,215.08) node [anchor=north west][inner sep=0.75pt]  [rotate=-359.68]  {$\boldsymbol{\tau} _{\theta z}$};
\draw (314.01,236.08) node [anchor=north west][inner sep=0.75pt]  [rotate=-359.68]  {$\boldsymbol{\tau} _{\theta z}$};
\draw (332.01,175.08) node [anchor=north west][inner sep=0.75pt]  [rotate=-359.68]  {$\boldsymbol{\tau} _{\theta z} +\delta \boldsymbol{\tau} _{\theta }$};
\draw (239.51,155.08) node [anchor=north west][inner sep=0.75pt]  [rotate=-359.68]  {$\boldsymbol{\tau} _{\theta z} +\delta \boldsymbol{\tau} _{\theta }$};
\draw (299,3.67) node [anchor=north west][inner sep=0.75pt]    {$\boldsymbol{u}$};
\draw (54.04,277.07) node [anchor=north west][inner sep=0.75pt]  [rotate=-318.72]  {$r$};
\draw (-4.58,252.95) node [anchor=north west][inner sep=0.75pt]  [rotate=-318.72]  {$z$};
\draw (254.34,274.98) node [anchor=north west][inner sep=0.75pt]  [rotate=-359.06]  {$\theta $};
\draw (222.22,225.04) node [anchor=north west][inner sep=0.75pt]  [rotate=-359.06]  {$z$};

\end{tikzpicture}

%% file: figures/noise_model.tex
\tikzset{every picture/.style={line width=0.75pt}} 

\begin{tikzpicture}[x=0.75pt,y=0.75pt,yscale=-1,xscale=1]

\draw [color={rgb, 255:red, 80; green, 227; blue, 194 }  ,draw opacity=1 ][line width=6]    (73.91,72.68) .. controls (104.58,73.68) and (101.58,107.18) .. (164.91,107.01) ;
\draw [color={rgb, 255:red, 80; green, 227; blue, 194 }  ,draw opacity=1 ][line width=6]    (163.91,107.01) .. controls (222.58,107.68) and (222.58,72.68) .. (250.58,72.68) ;

\draw [color={rgb, 255:red, 155; green, 155; blue, 155 }  ,draw opacity=1 ][line width=6]    (73.91,80.43) .. controls (104.58,81.43) and (101.58,114.93) .. (164.91,114.76) ;
\draw [color={rgb, 255:red, 155; green, 155; blue, 155 }  ,draw opacity=1 ][line width=6]    (163.91,114.76) .. controls (222.58,115.43) and (222.58,80.43) .. (250.58,80.43) ;

\draw    (73.91,77.43) -- (70.91,80.43) ;
\draw    (73.91,79.43) -- (70.91,82.43) ;
\draw    (73.91,81.43) -- (70.91,84.43) ;
\draw    (73.91,83.43) -- (70.91,86.43) ;
\draw    (250.45,77.14) -- (253.54,80.05) ;
\draw    (250.51,79.14) -- (253.59,82.05) ;
\draw    (250.57,81.13) -- (253.65,84.05) ;
\draw    (250.62,83.13) -- (253.71,86.05) ;
\draw  [color={rgb, 255:red, 74; green, 144; blue, 226 }  ,draw opacity=1 ][fill={rgb, 255:red, 74; green, 144; blue, 226 }  ,fill opacity=1 ] (217.69,88.21) -- (220.85,85.75) -- (223.32,88.9) -- (220.16,91.37) -- cycle ;
\draw  [color={rgb, 255:red, 74; green, 144; blue, 226 }  ,draw opacity=1 ][fill={rgb, 255:red, 74; green, 144; blue, 226 }  ,fill opacity=1 ] (109.65,89.55) -- (106.49,87.08) -- (104.02,90.24) -- (107.18,92.71) -- cycle ;
\draw [line width=0.75]  [dash pattern={on 3.75pt off 3pt on 7.5pt off 1.5pt}]  (163.58,57.26) -- (163.58,149.59) ;
\draw [shift={(163.58,152.59)}, rotate = 270] [fill={rgb, 255:red, 0; green, 0; blue, 0 }  ][line width=0.08]  [draw opacity=0] (7.14,-3.43) -- (0,0) -- (7.14,3.43) -- (4.74,0) -- cycle    ;
\draw [shift={(163.58,54.26)}, rotate = 90] [fill={rgb, 255:red, 0; green, 0; blue, 0 }  ][line width=0.08]  [draw opacity=0] (7.14,-3.43) -- (0,0) -- (7.14,3.43) -- (4.74,0) -- cycle    ;
\draw   (307.17,38.33) -- (359.5,38.33) -- (359.5,39.5) -- (307.17,39.5) -- cycle ;
\draw    (326,38.48) -- (318.86,31.34) ;
\draw    (321.24,38.48) -- (314.1,31.34) ;
\draw    (316.48,38.48) -- (309.35,31.34) ;
\draw    (311.72,38.48) -- (304.59,31.34) ;
\draw    (344,37.98) -- (336.86,30.84) ;
\draw    (339.24,37.98) -- (332.1,30.84) ;
\draw    (334.48,37.98) -- (327.35,30.84) ;
\draw    (329.72,37.98) -- (322.59,30.84) ;
\draw    (358.74,37.98) -- (351.6,30.84) ;
\draw    (353.98,37.98) -- (346.85,30.84) ;
\draw    (349.22,37.98) -- (342.09,30.84) ;
\draw   (346,37.98) -- (346,52.38) .. controls (350.88,52.58) and (355.05,54.58) .. (356.51,57.41) .. controls (357.97,60.24) and (356.42,63.32) .. (352.6,65.18) .. controls (349.63,66.61) and (345.78,67.19) .. (342.04,66.78) .. controls (340.58,66.78) and (339.4,66.06) .. (339.4,65.18) .. controls (339.4,64.29) and (340.58,63.58) .. (342.04,63.58) .. controls (345.78,63.16) and (349.63,63.75) .. (352.6,65.18) .. controls (355.78,66.84) and (357.57,69.15) .. (357.57,71.58) .. controls (357.57,74) and (355.78,76.31) .. (352.6,77.98) .. controls (349.63,79.41) and (345.78,79.99) .. (342.04,79.58) .. controls (340.58,79.58) and (339.4,78.86) .. (339.4,77.98) .. controls (339.4,77.09) and (340.58,76.38) .. (342.04,76.38) .. controls (345.78,75.96) and (349.63,76.55) .. (352.6,77.98) .. controls (355.78,79.64) and (357.57,81.95) .. (357.57,84.38) .. controls (357.57,86.8) and (355.78,89.11) .. (352.6,90.78) .. controls (349.63,92.21) and (345.78,92.79) .. (342.04,92.38) .. controls (340.58,92.38) and (339.4,91.66) .. (339.4,90.78) .. controls (339.4,89.89) and (340.58,89.18) .. (342.04,89.18) .. controls (345.78,88.76) and (349.63,89.35) .. (352.6,90.78) .. controls (356.42,92.63) and (357.97,95.71) .. (356.51,98.54) .. controls (355.05,101.37) and (350.88,103.37) .. (346,103.58) -- (346,117.98) ;
\draw   (317.24,38.48) -- (317.24,74.48) (327.68,82.48) -- (306.8,82.48) (327.68,74.48) -- (306.8,74.48) (317.24,82.48) -- (317.24,118.48) ;
\draw  [color={rgb, 255:red, 155; green, 155; blue, 155 }  ,draw opacity=1 ][fill={rgb, 255:red, 155; green, 155; blue, 155 }  ,fill opacity=1 ] (305.17,118.5) -- (359,118.5) -- (359,152) -- (305.17,152) -- cycle ;
\draw [line width=0.75]  [dash pattern={on 3.75pt off 3pt on 7.5pt off 1.5pt}]  (370,110.13) -- (370,161.87) ;
\draw [shift={(370,164.87)}, rotate = 270] [fill={rgb, 255:red, 0; green, 0; blue, 0 }  ][line width=0.08]  [draw opacity=0] (7.14,-3.43) -- (0,0) -- (7.14,3.43) -- (4.74,0) -- cycle    ;
\draw [shift={(370,107.13)}, rotate = 90] [fill={rgb, 255:red, 0; green, 0; blue, 0 }  ][line width=0.08]  [draw opacity=0] (7.14,-3.43) -- (0,0) -- (7.14,3.43) -- (4.74,0) -- cycle    ;
\draw  [dash pattern={on 4.5pt off 4.5pt}]  (295.5,136) -- (378,136) ;
\draw    (332.5,178.5) -- (332.5,156.5) ;
\draw [shift={(332.5,154.5)}, rotate = 90] [color={rgb, 255:red, 0; green, 0; blue, 0 }  ][line width=0.75]    (10.93,-3.29) .. controls (6.95,-1.4) and (3.31,-0.3) .. (0,0) .. controls (3.31,0.3) and (6.95,1.4) .. (10.93,3.29)   ;
\draw [color={rgb, 255:red, 80; green, 227; blue, 194 }  ,draw opacity=1 ][line width=6]    (73.91,65.18) .. controls (104.58,66.18) and (101.58,99.68) .. (164.91,99.51) ;
\draw [color={rgb, 255:red, 80; green, 227; blue, 194 }  ,draw opacity=1 ][line width=6]    (163.91,99.51) .. controls (222.58,100.18) and (222.58,65.18) .. (250.58,65.18) ;

\draw (147.41,52.26) node [anchor=north west][inner sep=0.75pt]    {$z$};
\draw (150.58,192.93) node [anchor=north west][inner sep=0.75pt]   [align=left] {(a)};
\draw (323.58,193.43) node [anchor=north west][inner sep=0.75pt]   [align=left] {(b)};
\draw (374,106.63) node [anchor=north west][inner sep=0.75pt]    {$z$};
\draw (363,66.13) node [anchor=north west][inner sep=0.75pt]    {$K$};
\draw (286,65.63) node [anchor=north west][inner sep=0.75pt]    {$C$};
\draw (270.5,110.13) node [anchor=north west][inner sep=0.75pt]    {$m_{\text{eff}}$};
\draw (255,155.5) node [anchor=north west][inner sep=0.75pt]  [font=\small] [align=left] {\begin{minipage}[lt]{51.38pt}\setlength\topsep{0pt}
\begin{center}
{\small Thermal}\\{\small displacement}
\end{center}

\end{minipage}};

\end{tikzpicture}

%% file: figures/Wouter_geometry.pdf_tex
\begingroup%
  \makeatletter%
  \providecommand\color[2][]{%
    \errmessage{(Inkscape) Color is used for the text in Inkscape, but the package 'color.sty' is not loaded}%
    \renewcommand\color[2][]{}%
  }%
  \providecommand\transparent[1]{%
    \errmessage{(Inkscape) Transparency is used (non-zero) for the text in Inkscape, but the package 'transparent.sty' is not loaded}%
    \renewcommand\transparent[1]{}%
  }%
  \providecommand\rotatebox[2]{#2}%
  \newcommand*\fsize{\dimexpr\f@size pt\relax}%
  \newcommand*\lineheight[1]{\fontsize{\fsize}{#1\fsize}\selectfont}%
  \ifx\svgwidth\undefined%
    \setlength{\unitlength}{665.08872986bp}%
    \ifx\svgscale\undefined%
      \relax%
    \else%
      \setlength{\unitlength}{\unitlength * \real{\svgscale}}%
    \fi%
  \else%
    \setlength{\unitlength}{\svgwidth}%
  \fi%
  \global\let\svgwidth\undefined%
  \global\let\svgscale\undefined%
  \makeatother%
  \begin{picture}(1,0.28442409)%
    \lineheight{1}%
    \setlength\tabcolsep{0pt}%
    \put(0,0){\includegraphics[width=\unitlength,page=1]{Wouter_geometry.pdf}}%
    \put(0.17986449,0.10859218){\makebox(0,0)[lt]{\lineheight{5}\smash{\begin{tabular}[t]{l}$r_{WG}^1$\end{tabular}}}}%
    \put(0.26581107,0.14820232){\makebox(0,0)[lt]{\lineheight{5}\smash{\begin{tabular}[t]{l}$r_{WG}^2$\end{tabular}}}}%
    \put(0,0){\includegraphics[width=\unitlength,page=2]{Wouter_geometry.pdf}}%
    \put(0.7290486,0.12224664){\makebox(0,0)[lt]{\lineheight{5}\smash{\begin{tabular}[t]{l}$d_M$\end{tabular}}}}%
    \put(0.25829047,0.00620367){\makebox(0,0)[lt]{\lineheight{5}\smash{\begin{tabular}[t]{l}(a)\end{tabular}}}}%
    \put(0.1109709,0.2699951){\makebox(0,0)[lt]{\lineheight{5}\smash{\begin{tabular}[t]{l}Racetrack resonator\end{tabular}}}}%
    \put(0.73256949,0.00617649){\makebox(0,0)[lt]{\lineheight{5}\smash{\begin{tabular}[t]{l}(b)\end{tabular}}}}%
    \put(0.96454707,0.20425446){\makebox(0,0)[lt]{\lineheight{5}\smash{\begin{tabular}[t]{l}$h_M$\end{tabular}}}}%
    \put(0.96679653,0.09738176){\makebox(0,0)[lt]{\lineheight{5}\smash{\begin{tabular}[t]{l}$h_S$\end{tabular}}}}%
    \put(0.83002206,0.27002227){\makebox(0,0)[lt]{\lineheight{5}\smash{\begin{tabular}[t]{l}$h_{WG}$\end{tabular}}}}%
    \put(0.75350419,0.26180129){\makebox(0,0)[lt]{\lineheight{5}\smash{\begin{tabular}[t]{l}$r_{WG}^1$\end{tabular}}}}%
    \put(0.59770319,0.26254867){\makebox(0,0)[lt]{\lineheight{5}\smash{\begin{tabular}[t]{l}$w_{WG}$\end{tabular}}}}%
    \put(0,0){\includegraphics[width=\unitlength,page=3]{Wouter_geometry.pdf}}%
    \put(0.34945967,0.2653701){\makebox(0,0)[lt]{\lineheight{5}\smash{\begin{tabular}[t]{l}Pass port\end{tabular}}}}%
    \put(0,0){\includegraphics[width=\unitlength,page=4]{Wouter_geometry.pdf}}%
    \put(0.00445964,0.09273908){\makebox(0,0)[lt]{\lineheight{5}\smash{\begin{tabular}[t]{l}Light in\end{tabular}}}}%
    \put(0,0){\includegraphics[width=\unitlength,page=5]{Wouter_geometry.pdf}}%
    \put(0.38710879,0.05686573){\makebox(0,0)[lt]{\lineheight{5}\smash{\begin{tabular}[t]{l}Drop port\end{tabular}}}}%
    \put(0,0){\includegraphics[width=\unitlength,page=6]{Wouter_geometry.pdf}}%
  \end{picture}%
\endgroup%

%% file: figures/disp_time.tex
\pgfplotsset{cycle list/Dark2}
\begin{tikzpicture}
    \begin{axis}[
    scale only axis, 
    scaled x ticks=false,
    scaled y ticks=false, 
    width=5.25cm,
    height=5cm,
    legend style={draw=none, at={(0.32,0.3)}, legend cell align={left}, fill=none, anchor=north},
	xmin=5e-5,
	xmax=1e-4,
	xtick distance=1e-5,
	ytick distance=2e-8,
	ymin=-8e-8,
	ymax=8e-8,
	ytick={-8e-8,-6e-8,-4e-8,-2e-8,0,2e-8,4e-8,6e-8,8e-8},
	yticklabels={-80,,,,0,,,,80},
	xtick={5e-5,6e-5,7e-5,8e-5,9e-5,1e-4},
 scaled x ticks=false,
	xticklabels={0,,,,,100},
	xlabel={Time (\SI{}{\micro\second})},
	ylabel={Axial displacement (\SI{}{\nano\meter})}, mark repeat={15},
	]
    

    \addplot+ [color=teal,thick,smooth] file{figures/wouter_disp_time.dat};
    \end{axis}
    \end{tikzpicture}

%% file: figures/disp_freq.tex
\pgfplotsset{cycle list/Dark2}
\begin{tikzpicture}
    \begin{axis}[
    scale only axis, 
    width=5.25cm,
    height=5cm,
    legend style={draw=none, at={(0.32,0.3)}, legend cell align={left}, fill=none, anchor=north},
	xmin=0e-6,
	xmax=1.5e6,
	xtick distance=3e5,
	ytick distance=1e-9,
	ymin=0,
	ymax=1e-8,
	ytick={0,1e-9,2e-9,3e-9,4e-9,5e-9,6e-9,7e-9,8e-9,8e-9,1e-8},
	yticklabels={0,,,,,,,,,,10},
	xtick={0,3e5,6e5,9e5,1.2e6,1.5e6},
 scaled x ticks=false,
 scaled y ticks=false,
	xticklabels={0,,,,,1.5},
	xlabel={Frequency (\SI{}{\mega\hertz})},
	ylabel={Axial displacement amplitude}, mark repeat={15},
 extra x ticks = {5.3e5,6.15e5,7e5},
	extra x tick labels = {$f_1$,$f_c$,$f_2$},
	]
    
    \draw [gray, fill=gray, opacity= 0.4, draw opacity =0] (axis cs:5.9e5,0)  rectangle (axis cs:6.42e5,1e-8);

    \addplot+ [color=red,thick,smooth] file{figures/wouter_disp_freq.dat};
    \end{axis}
    \end{tikzpicture}

%% file: figures/radial_disp_.tex
\pgfplotsset{cycle list/Dark2}
\begin{tikzpicture}
    \begin{axis}[
    scale only axis, 
    width=6cm,
    height=5cm,
    legend style={draw=none, at={(0.32,0.3)}, legend cell align={left}, fill=none, anchor=north},
	xmin=0,
	xmax=2.2e-7,
	xtick distance=2e-8,
	ytick distance=0.5e-11,
	ymin=-10e-11,
	ymax=-7e-11,
	ytick={-10e-11,-9.5e-11,-9e-11,-8.5e-11,-8e-11,-7.5e-11,-7e-11},
	yticklabels={-0.1,,,,,,-0.07},
	xtick={0,2e-8,4e-8,6e-8,8e-8,1e-7,1.2e-7,1.4e-7,1.6e-7,1.8e-7,2e-7,2.2e-7},
 scaled x ticks=false,
 scaled y ticks=false,
	xticklabels={0,,,,,,,,,,,0.22},
	xlabel={Thickness of the waveguide (\SI{}{\micro\meter})},
	ylabel={Radial displacement (\SI{}{\nano\meter})}, mark repeat={15},
	]
    \node [above right] at (rel axis cs:0.6,0.2) {\includegraphics[scale=0.25]{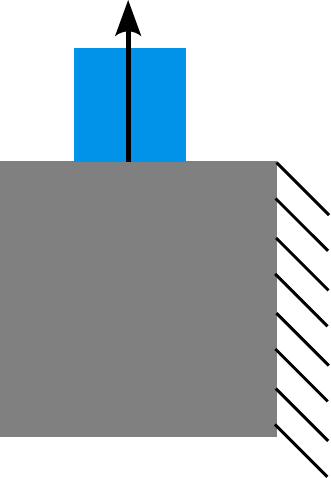}};

    \addplot+ [color=red,thick,smooth] file{figures/wouter_radial_disp_freq.dat};
    \end{axis}
    \end{tikzpicture}

%% file: figures/coordinate_transform.tex
  

\begin{tikzpicture}[x=0.75pt,y=0.75pt,yscale=-1,xscale=1]

\draw  [fill={rgb, 255:red, 193; green, 193; blue, 193 }  ,fill opacity=1 ] (226.5,108) -- (371,108) -- (371,129.5) -- (226.5,129.5) -- cycle ;
\draw  [fill={rgb, 255:red, 193; green, 193; blue, 193 }  ,fill opacity=1 ] (226.5,108) -- (226.5,129.5) -- (191.1,168.9) -- (191.1,146.4) -- (191.1,146.4) -- cycle ;
\draw  [draw opacity=0] (189.67,4.03) .. controls (189.71,4.03) and (189.76,4.03) .. (189.8,4.03) .. controls (289.39,3.78) and (370.3,50.25) .. (371,107.93) -- (190.06,108.98) -- cycle ; \draw   (189.67,4.03) .. controls (189.71,4.03) and (189.76,4.03) .. (189.8,4.03) .. controls (289.39,3.78) and (370.3,50.25) .. (371,107.93) ;  
\draw  [fill={rgb, 255:red, 74; green, 144; blue, 226 }  ,fill opacity=1 ] (311.1,92.5) -- (328.5,92.5) -- (328.5,107.9) -- (311.1,107.9) -- cycle ;
\draw  [draw opacity=0] (190.59,19.04) .. controls (266.6,19.55) and (327.93,52.13) .. (328.5,92.5) -- (188.3,93.31) -- cycle ; \draw   (190.59,19.04) .. controls (266.6,19.55) and (327.93,52.13) .. (328.5,92.5) ;  
\draw  [draw opacity=0] (191.12,28.11) .. controls (191.34,28.11) and (191.56,28.11) .. (191.77,28.11) .. controls (257.35,27.95) and (310.63,57.06) .. (311.11,93.21) -- (191.94,93.89) -- cycle ; \draw   (191.12,28.11) .. controls (191.34,28.11) and (191.56,28.11) .. (191.77,28.11) .. controls (257.35,27.95) and (310.63,57.06) .. (311.11,93.21) ;  
\draw  [fill={rgb, 255:red, 193; green, 193; blue, 193 }  ,fill opacity=1 ] (238.1,212) -- (333.32,212) -- (333.32,237) -- (238.1,237) -- cycle ;
\draw  [fill={rgb, 255:red, 74; green, 144; blue, 226 }  ,fill opacity=1 ] (274.72,193.4) -- (294.95,193.4) -- (294.95,211.31) -- (274.72,211.31) -- cycle ;
\draw    (294.95,193.4) -- (319.6,180.4) ;
\draw    (294.95,211.31) -- (319.6,198.31) ;
\draw    (333.32,212) -- (357.97,199) ;
\draw    (333.32,237) -- (357.97,224) ;
\draw    (238.1,212) -- (262.75,199) ;
\draw    (319.6,198.31) -- (357.97,198.31) ;
\draw    (274.72,193.4) -- (299.37,180.4) ;
\draw    (299.37,180.4) -- (319.6,180.4) ;
\draw    (262.75,199) -- (274.6,199) ;
\draw    (319.6,180.4) -- (319.6,198.31) ;
\draw    (357.97,199) -- (357.97,224) ;

\draw  [dash pattern={on 3.75pt off 3pt on 7.5pt off 1.5pt}]  (226.5,129.5) -- (226.5,74.4) ;
\draw [shift={(226.5,71.4)}, rotate = 90] [fill={rgb, 255:red, 0; green, 0; blue, 0 }  ][line width=0.08]  [draw opacity=0] (10.72,-5.15) -- (0,0) -- (10.72,5.15) -- (7.12,0) -- cycle    ;
\draw  [dash pattern={on 3.75pt off 3pt on 7.5pt off 1.5pt}]  (226.5,108) -- (287.74,88.32) ;
\draw [shift={(290.6,87.4)}, rotate = 162.18] [fill={rgb, 255:red, 0; green, 0; blue, 0 }  ][line width=0.08]  [draw opacity=0] (10.72,-5.15) -- (0,0) -- (10.72,5.15) -- (7.12,0) -- cycle    ;
\draw  [draw opacity=0][dash pattern={on 3.75pt off 3pt on 7.5pt off 1.5pt}] (256.04,81.27) .. controls (256.36,81.41) and (256.68,81.57) .. (256.99,81.73) .. controls (266.88,86.74) and (271.89,97.56) .. (269.98,107.93) -- (245.96,103.51) -- cycle ; \draw  [dash pattern={on 3.75pt off 3pt on 7.5pt off 1.5pt}] (256.04,81.27) .. controls (256.36,81.41) and (256.68,81.57) .. (256.99,81.73) .. controls (266.88,86.74) and (271.89,97.56) .. (269.98,107.93) ;  
\draw    (256.04,81.27) -- (251.19,78.42) ;
\draw [shift={(248.6,76.9)}, rotate = 30.42] [fill={rgb, 255:red, 0; green, 0; blue, 0 }  ][line width=0.08]  [draw opacity=0] (10.72,-5.15) -- (0,0) -- (10.72,5.15) -- (7.12,0) -- cycle    ;

\draw  [dash pattern={on 3.75pt off 3pt on 7.5pt off 1.5pt}]  (196.1,261.4) -- (196.1,229.4) ;
\draw [shift={(196.1,226.4)}, rotate = 90] [fill={rgb, 255:red, 0; green, 0; blue, 0 }  ][line width=0.08]  [draw opacity=0] (10.72,-5.15) -- (0,0) -- (10.72,5.15) -- (7.12,0) -- cycle    ;
\draw  [dash pattern={on 3.75pt off 3pt on 7.5pt off 1.5pt}]  (196.1,261.4) -- (221.6,261.4) ;
\draw [shift={(224.6,261.4)}, rotate = 180] [fill={rgb, 255:red, 0; green, 0; blue, 0 }  ][line width=0.08]  [draw opacity=0] (10.72,-5.15) -- (0,0) -- (10.72,5.15) -- (7.12,0) -- cycle    ;
\draw  [dash pattern={on 3.75pt off 3pt on 7.5pt off 1.5pt}]  (196.1,261.4) -- (214.98,242.52) ;
\draw [shift={(217.1,240.4)}, rotate = 135] [fill={rgb, 255:red, 0; green, 0; blue, 0 }  ][line width=0.08]  [draw opacity=0] (10.72,-5.15) -- (0,0) -- (10.72,5.15) -- (7.12,0) -- cycle    ;
\draw  [fill={rgb, 255:red, 0; green, 0; blue, 0 }  ,fill opacity=1 ] (295,154.57) -- (300.06,154.57) -- (300.06,139.2) -- (303.51,139.2) -- (303.51,154.57) -- (308.58,154.57) -- (301.79,163.2) -- cycle ;

\draw (274,67.5) node [anchor=north west][inner sep=0.75pt]    {$r$};
\draw (254,57.5) node [anchor=north west][inner sep=0.75pt]    {$\theta $};
\draw (208,65) node [anchor=north west][inner sep=0.75pt]    {$z$};
\draw (215,228) node [anchor=north west][inner sep=0.75pt]    {$y$};
\draw (224.5,255) node [anchor=north west][inner sep=0.75pt]    {$x$};
\draw (191.1,213) node [anchor=north west][inner sep=0.75pt]    {$z$};

\end{tikzpicture}

%% file: figures/stresses.tex
\pgfplotsset{cycle list/Dark2}
\begin{tikzpicture}
    \begin{axis}[
    scale only axis, 
    width=6cm,
    height=5cm,
    legend style={draw=none, at={(0.22,0.85)}, legend cell align={left}, fill=none, anchor=north},
	xmin=0,
	xmax=2.2e-7,
	xtick distance=2e-8,
	ytick distance=2e5,
	ymin=-1e6,
	ymax=2e5,
	ytick={-1e6,-8e5,-6e5,-4e5,-2e5,0,2e5},
	yticklabels={-1,,,,,0,0.2},
	xtick={0,2e-8,4e-8,6e-8,8e-8,1e-7,1.2e-7,1.4e-7,1.6e-7,1.8e-7,2e-7,2.2e-7},
 scaled x ticks=false,
 scaled y ticks=false,
	xticklabels={0,,,,,,,,,,,0.22},
	xlabel={Thickness of the waveguide (\SI{}{\micro\meter})},
	ylabel={Stresses (\SI{}{\mega\pascal})}, mark repeat={15},
	]

    \addplot+ [color=blue,smooth] file{figures/wouter_radial_stress_freq.dat};
    \addplot+ [color=teal,thick,smooth, dotted] file{figures/wouter_angular_stress_freq.dat};
    \addplot+ [color=red,thick,smooth, dashed] file{figures/wouter_axial_stress_freq.dat};
    \legend{$\sigma_r (x)$, $\sigma_\theta (y)$, $\sigma_z (z)$}
    \end{axis}
    \end{tikzpicture}

%% file: figures/IPUT_3D_model.pdf_tex
\begingroup%
  \makeatletter%
  \providecommand\color[2][]{%
    \errmessage{(Inkscape) Color is used for the text in Inkscape, but the package 'color.sty' is not loaded}%
    \renewcommand\color[2][]{}%
  }%
  \providecommand\transparent[1]{%
    \errmessage{(Inkscape) Transparency is used (non-zero) for the text in Inkscape, but the package 'transparent.sty' is not loaded}%
    \renewcommand\transparent[1]{}%
  }%
  \providecommand\rotatebox[2]{#2}%
  \newcommand*\fsize{\dimexpr\f@size pt\relax}%
  \newcommand*\lineheight[1]{\fontsize{\fsize}{#1\fsize}\selectfont}%
  \ifx\svgwidth\undefined%
    \setlength{\unitlength}{279.68358318bp}%
    \ifx\svgscale\undefined%
      \relax%
    \else%
      \setlength{\unitlength}{\unitlength * \real{\svgscale}}%
    \fi%
  \else%
    \setlength{\unitlength}{\svgwidth}%
  \fi%
  \global\let\svgwidth\undefined%
  \global\let\svgscale\undefined%
  \makeatother%
  \begin{picture}(1,1.66832622)%
    \lineheight{1}%
    \setlength\tabcolsep{0pt}%
    \put(0,0){\includegraphics[width=\unitlength,page=1]{IPUT_3D_model.pdf}}%
    \put(0.23181791,1.63592125){\makebox(0,0)[lt]{\lineheight{5}\smash{\begin{tabular}[t]{l}$p_{in}$\end{tabular}}}}%
    \put(0,0){\includegraphics[width=\unitlength,page=2]{IPUT_3D_model.pdf}}%
    \put(0.73715831,1.58855534){\makebox(0,0)[lt]{\lineheight{5}\smash{\begin{tabular}[t]{l}Symmetric BC\end{tabular}}}}%
    \put(0,0){\includegraphics[width=\unitlength,page=3]{IPUT_3D_model.pdf}}%
    \put(0.04231761,0.7806864){\makebox(0,0)[lt]{\lineheight{5}\smash{\begin{tabular}[t]{l}Symmetric BC\end{tabular}}}}%
    \put(0,0){\includegraphics[width=\unitlength,page=4]{IPUT_3D_model.pdf}}%
    \put(0.27517223,0.05442354){\makebox(0,0)[lt]{\lineheight{5}\smash{\begin{tabular}[t]{l}Symmetric BC\end{tabular}}}}%
    \put(0,0){\includegraphics[width=\unitlength,page=5]{IPUT_3D_model.pdf}}%
    \put(0.77902179,0.68556006){\makebox(0,0)[lt]{\lineheight{5}\smash{\begin{tabular}[t]{l}Fixed BC\end{tabular}}}}%
    \put(0.61663044,1.30734474){\makebox(0,0)[lt]{\lineheight{5}\smash{\begin{tabular}[t]{l}$w_C$\end{tabular}}}}%
    \put(0.83094994,1.18890817){\makebox(0,0)[lt]{\lineheight{5}\smash{\begin{tabular}[t]{l}$h_C$\end{tabular}}}}%
    \put(0.1809939,0.07821006){\makebox(0,0)[lt]{\lineheight{5}\smash{\begin{tabular}[t]{l}$x$\end{tabular}}}}%
    \put(0.04640967,0.03110315){\makebox(0,0)[lt]{\lineheight{5}\smash{\begin{tabular}[t]{l}$y$\end{tabular}}}}%
    \put(0.13450899,0.13829077){\makebox(0,0)[lt]{\lineheight{5}\smash{\begin{tabular}[t]{l}$z$\end{tabular}}}}%
    \put(0,0){\includegraphics[width=\unitlength,page=6]{IPUT_3D_model.pdf}}%
  \end{picture}%
\endgroup%

%% file: figures/radial_disp_3D.tex
\pgfplotsset{cycle list/Dark2}
\begin{tikzpicture}
    \begin{axis}[
    scale only axis, 
   width=5.25cm,
    height=5cm,
    legend style={draw=none, at={(0.32,0.3)}, legend cell align={left}, fill=none, anchor=north},
	xmin=0,
	xmax=2.8e-5,
	xtick distance=2e-6,
	ytick distance=5e-11,
	ymin=-2e-10,
	ymax=-4e-11,
	ytick={-2e-10,-1.6e-10,-1.2e-10,-8e-11,-4e-11},
	yticklabels={-0.2,,,,-0.04},
	xtick={0,2e-6,4e-6,6e-6,8e-6,1e-5,1.2e-5,1.4e-5,1.6e-5,1.8e-5,2e-5,2.2e-5,2.4e-5,2.6e-5,2.8e-5},
 scaled x ticks=false,
 scaled y ticks=false,
	xticklabels={0,,,,,,,,,,,,,,28},
	xlabel={Distance along the waveguide (\SI{}{\micro\meter})},
	ylabel={Radial displacement (\SI{}{\nano\meter})}, mark repeat={15},
	]
    \node [above right] at (rel axis cs:0.1,0.2) {\includegraphics[scale=0.35]{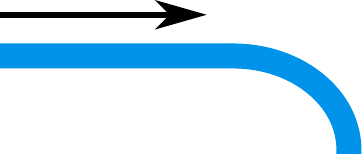}};

    \addplot+ [color=red,thick,smooth] file{figures/wouter_radial_disp_freq_3D.dat};
    \end{axis}
    \end{tikzpicture}

%% file: figures/stresses_3D.tex
\pgfplotsset{cycle list/Dark2}
\begin{tikzpicture}
    \begin{axis}[
    scale only axis, 
    width=5.25cm,
    height=5cm,
    legend style={draw=none, at={(0.22,0.85)}, legend cell align={left}, fill=none, anchor=north},
	xmin=0,
	xmax=2.2e-7,
	xtick distance=2e-8,
	ytick distance=2e5,
	ymin=-1e6,
	ymax=2e5,
	ytick={-1e6,-8e5,-6e5,-4e5,-2e5,0,2e5},
	yticklabels={-1,,,,,,0.2},
	xtick={0,2e-8,4e-8,6e-8,8e-8,1e-7,1.2e-7,1.4e-7,1.6e-7,1.8e-7,2e-7,2.2e-7},
 scaled x ticks=false,
 scaled y ticks=false,
	xticklabels={0,,,,,,,,,,,0.22},
	xlabel={Thickness of the waveguide (\SI{}{\micro\meter})},
	ylabel={Stresses (\SI{}{\mega\pascal})}, mark repeat={15},
	]
 \node [above right] at (rel axis cs:0.6,0.4) {\includegraphics[scale=0.2]{figures/WG_membrane_.pdf}};

    \addplot+ [color=blue,smooth] file{figures/wouter_stress_x.dat};
    \addplot+ [color=teal,thick,smooth, dotted] file{figures/wouter_stress_y.dat};
    \addplot+ [color=red,thick,smooth, dashed] file{figures/wouter_stress_z.dat};
    \legend{$\sigma_x$, $\sigma_y$, $\sigma_z$}
    \end{axis}
    \end{tikzpicture}

%% file: figures/NEP_freq.tex
\pgfplotsset{cycle list/Dark2}
\begin{tikzpicture}
    \begin{semilogyaxis}[
    scale only axis, 
    width=5.25cm,
    height=5cm,
    legend style={draw=none, at={(0.72,0.3)}, legend cell align={left}, fill=none, anchor=north},
	xmin=1e5,
	xmax=5e6,
	xtick distance=5e5,
	ytick distance=0.025,
	ymin=0.005,
	ymax=0.5,
	ytick={0.005,0.05,0.5},
	yticklabels={0.005,0.05,0.5},
	xtick={1e5,6e5,1.1e6,1.6e6,2.1e6,2.6e6,3.1e6,3.6e6,4.1e6,4.6e6,5e6},
 scaled x ticks=false,
 scaled y ticks=false,
	xticklabels={0.1,,,,,,,,,,5},
	xlabel={Frequency (\SI{}{\mega\hertz})},
	ylabel={NEP (\SI{}{\pascal})}, mark repeat={15},
	]
 \addplot[color=red,
   only marks,
   mark=triangle,
   ] coordinates {
    (0.76e6,0.4)
    };

    \addplot+ [color=teal,thick,smooth] file{figures/NEP_freq.dat};
    \legend{NEP\textsubscript{expt}, NEP\textsubscript{thermal}}
    \end{semilogyaxis}
    \end{tikzpicture}

%% file: figures/NEP_Q.tex
\pgfplotsset{cycle list/Dark2}
\begin{tikzpicture}
    \begin{semilogyaxis}[
    scale only axis, 
    width=5.25cm,
    height=5cm,
    legend style={draw=none, at={(0.72,0.3)}, legend cell align={left}, fill=none, anchor=north},
	xmin=1,
	xmax=15,
	xtick distance=1,
	ytick distance=0.025,
	ymin=0.005,
	ymax=0.5,
	ytick={0.005,0.05,0.5},
	yticklabels={0.005,0.05,0.5},
	xtick={1,2,3,4,5,6,7,8,9,10,11,12,13,14,15},
 scaled x ticks=false,
 scaled y ticks=false,
	xticklabels={1,,,,,,,,,,,,,,15},
	xlabel={Q factor},
	ylabel={NEP (\SI{}{\pascal})}, mark repeat={15},
	]
 \addplot[color=red,
   only marks,
   mark=triangle,
   ] coordinates {
    (10,0.4)
    };

    \addplot+ [color=teal,thick,smooth] file{figures/NEP_Q.dat};
    \legend{NEP\textsubscript{expt}, NEP\textsubscript{thermal}}
    \end{semilogyaxis}
    \end{tikzpicture}

%% file: cas-dc-template.bbl
\begin{thebibliography}{10}

\bibitem{Carovac2011-sz}
Aladin Carovac, Fahrudin Smajlovic, and Dzelaludin Junuzovic.
\newblock Application of ultrasound in medicine.
\newblock {\em Acta Inform. Med.}, 19(3):168--171, September 2011.

\bibitem{DRINKWATER2006525}
Bruce~W. Drinkwater and Paul~D. Wilcox.
\newblock Ultrasonic arrays for non-destructive evaluation: A review.
\newblock {\em NDT \& E International}, 39(7):525--541, 2006.

\bibitem{SANDERSON2002125}
M.L Sanderson and H~Yeung.
\newblock Guidelines for the use of ultrasonic non-invasive metering
  techniques.
\newblock {\em Flow Measurement and Instrumentation}, 13(4):125--142, 2002.

\bibitem{Carullo2001AnUS}
Alessio Carullo and Marco Parvis.
\newblock An ultrasonic sensor for distance measurement in automotive
  applications.
\newblock {\em IEEE Sensors Journal}, 1:143--, 2001.

\bibitem{AWAD2012410}
T.S. Awad, H.A. Moharram, O.E. Shaltout, D.~Asker, and M.M. Youssef.
\newblock Applications of ultrasound in analysis, processing and quality
  control of food: A review.
\newblock {\em Food Research International}, 48(2):410--427, 2012.

\bibitem{Insight_2022}
Global~Market Insight.
\newblock Diagnostic ultrasound market share: Statistics report, 2023-2032,
  2022.

\bibitem{Markets_2023}
Markets~and Markets.
\newblock Ultrasound market by technology, global forcast to 2028, 2023.

\bibitem{Tressler1998}
James~F. Tressler, Sedat Alkoy, and Robert~E. Newnham.
\newblock Piezoelectric sensors and sensor materials.
\newblock {\em Journal of Electroceramics}, 2(4):257--272, Dec 1998.

\bibitem{Ilkhechi:20}
Afshin~Kashani Ilkhechi, Christopher Ceroici, Zhenhao Li, and Roger Zemp.
\newblock Transparent capacitive micromachined ultrasonic transducer (cmut)
  arrays for real-time photoacoustic applications.
\newblock {\em Opt. Express}, 28(9):13750--13760, Apr 2020.

\bibitem{Jung_2017}
Joontaek Jung, Wonjun Lee, Woojin Kang, Eunjung Shin, Jungho Ryu, and Hongsoo
  Choi.
\newblock Review of piezoelectric micromachined ultrasonic transducers and
  their applications.
\newblock {\em Journal of Micromechanics and Microengineering}, 27(11):113001,
  sep 2017.

\bibitem{https://doi.org/10.1118/1.4792462}
Wenfeng Xia, Daniele Piras, Johan C.~G. van Hespen, Spiridon van Veldhoven,
  Christian Prins, Ton~G. van Leeuwen, Wiendelt Steenbergen, and Srirang
  Manohar.
\newblock An optimized ultrasound detector for photoacoustic breast tomography.
\newblock {\em Medical Physics}, 40(3):032901, 2013.

\bibitem{Manwar2020-vy}
Rayyan Manwar, Karl Kratkiewicz, and Kamran Avanaki.
\newblock Overview of ultrasound detection technologies for photoacoustic
  imaging.
\newblock {\em Micromachines (Basel)}, 11(7):692, July 2020.

\bibitem{Westerveld2021}
Wouter~J. Westerveld, Md. Mahmud-Ul-Hasan, Rami Shnaiderman, Vasilis
  Ntziachristos, Xavier Rottenberg, Simone Severi, and Veronique Rochus.
\newblock Sensitive, small, broadband and scalable optomechanical ultrasound
  sensor in silicon photonics.
\newblock {\em Nature Photonics}, 15(5):341--345, May 2021.

\bibitem{Ouyang:19}
Boling Ouyang, Yanlu Li, Marten Kruidhof, Roland Horsten, Koen W.~A. van
  Dongen, and Jacob Caro.
\newblock On-chip silicon mach\&\#x2013;zehnder interferometer sensor for
  ultrasound detection.
\newblock {\em Opt. Lett.}, 44(8):1928--1931, Apr 2019.

\bibitem{Leinders2015}
S.~M. Leinders, W.~J. Westerveld, J.~Pozo, P.~L. M.~J. van Neer, B.~Snyder,
  P.~O'Brien, H.~P. Urbach, N.~de~Jong, and M.~D. Verweij.
\newblock A sensitive optical micro-machined ultrasound sensor (omus) based on
  a silicon photonic ring resonator on an acoustical membrane.
\newblock {\em Scientific Reports}, 5(1):14328, Sep 2015.

\bibitem{Jansen2016MicroOptoMechanicalPS}
Roelof Jansen, Veronique Rochus, Jeroen Goyvaerts, Guy A.~E. Vandenbosch, Bob
  Voort, Pieter Neutens, John~M. O'Callaghan, Harrie A.~C. Tilmans, and Xavier
  Rottenberg.
\newblock Micro-opto-mechanical pressure sensor (momps) in sin integrated
  photonics platform.
\newblock 2016.

\bibitem{7926285}
V.~Rochus, R.~Jansen, R.~Haouari, B.~Figeys, V.~Mukund, F.~Verhaegen,
  J.~Goyvaerts, P.~Neutens, J.~O'~Callaghan, A.~Stassen, S.~Lenci, and
  X.~Rottenberg.
\newblock Modelling and design of micro-opto-mechanical pressure sensors in the
  presence of residual stresses.
\newblock In {\em 2017 18th International Conference on Thermal, Mechanical and
  Multi-Physics Simulation and Experiments in Microelectronics and Microsystems
  (EuroSimE)}, pages 1--5, 2017.

\bibitem{9152628}
Maja Zunic, Wouter~J. Westerveld, Pieter Gijsenbergh, Yongbin Jeong, Alessio
  Miranda, John O’Callaghan, Hamideh Jafarpoorchekab, Emmanuel~Vander
  Poorten, Xavier Rottenberg, Roelof Jansen, and Veronique Rochus.
\newblock Design of a micro-opto-mechanical ultrasound sensor for photoacoustic
  imaging.
\newblock In {\em 2020 21st International Conference on Thermal, Mechanical and
  Multi-Physics Simulation and Experiments in Microelectronics and Microsystems
  (EuroSimE)}, pages 1--8, 2020.

\bibitem{Rochus_2018}
V~Rochus, R~Jansen, J~Goyvaerts, P~Neutens, J~O’Callaghan, and X~Rottenberg.
\newblock Fast analytical model of mzi micro-opto-mechanical pressure sensor.
\newblock {\em Journal of Micromechanics and Microengineering}, 28(6):064003,
  mar 2018.

\bibitem{8724569}
V.~Rochus, W.J. Westerveld, B.~Figeys, X.~Rottenberg, and R.~Jansen.
\newblock Simulation and design of an optical accelerometer.
\newblock In {\em 2019 20th International Conference on Thermal, Mechanical and
  Multi-Physics Simulation and Experiments in Microelectronics and Microsystems
  (EuroSimE)}, pages 1--6, 2019.

\bibitem{8369932}
H.~Gao, C.~H. Huang, W.~Westerveld, R.~Haouari, B.~Troia, F.~Verhaegen,
  R.~Jansen, B.~Figeys, X.~Rottenberg, and V.~Rochus.
\newblock Simulation and characterization of a high-sensitive
  micro-opto-mechanical microphone.
\newblock In {\em 2018 19th International Conference on Thermal, Mechanical and
  Multi-Physics Simulation and Experiments in Microelectronics and Microsystems
  (EuroSimE)}, pages 1--4, 2018.

\bibitem{8724528}
W.J. Westerveld, S.M. Leinders, P.L.M.J. van Neer, H.P. Urbach, N.~de Jong,
  M.D. Verweij, X.~Rottenberg, and V.~Rochus.
\newblock Optical micro-machined ultrasound sensors with a silicon photonic
  resonator in a buckled acoustical membrane.
\newblock In {\em 2019 20th International Conference on Thermal, Mechanical and
  Multi-Physics Simulation and Experiments in Microelectronics and Microsystems
  (EuroSimE)}, pages 1--7, 2019.

\bibitem{ZHANG2017113}
Senlin Zhang, Jian Chen, and Sailing He.
\newblock Novel ultrasound detector based on small slot micro-ring resonator
  with ultrahigh q factor.
\newblock {\em Optics Communications}, 382:113--118, 2017.

\bibitem{Lee2023}
Youngseop Lee, Hao~F. Zhang, and Cheng Sun.
\newblock Highly sensitive ultrasound detection using nanofabricated polymer
  micro-ring resonators.
\newblock {\em Nano Convergence}, 10(1):30, Jun 2023.

\bibitem{Allen:11}
T.~W. Allen, J.~Silverstone, N.~Ponnampalam, T.~Olsen, A.~Meldrum, and R.~G.
  DeCorby.
\newblock High-finesse cavities fabricated by buckling self-assembly of
  a-si/sio2 multilayers.
\newblock {\em Opt. Express}, 19(20):18903--18909, Sep 2011.

\bibitem{Bitarafan:15}
M.~H. Bitarafan, H.~Ramp, T.~W. Allen, C.~Potts, X.~Rojas, A.~J.~R. MacDonald,
  J.~P. Davis, and R.~G. DeCorby.
\newblock Thermomechanical characterization of on-chip buckled dome
  fabry\&\#x2013;perot microcavities.
\newblock {\em J. Opt. Soc. Am. B}, 32(6):1214--1220, Jun 2015.

\bibitem{PhysRevD.42.2437}
Peter~R. Saulson.
\newblock Thermal noise in mechanical experiments.
\newblock {\em Phys. Rev. D}, 42:2437--2445, Oct 1990.

\bibitem{RevModPhys.68.801}
Charles~H. Henry and Rudolf~F. Kazarinov.
\newblock Quantum noise in photonics.
\newblock {\em Rev. Mod. Phys.}, 68:801--853, Jul 1996.

\bibitem{doi:10.1126/science.1156032}
T.~J. Kippenberg and K.~J. Vahala.
\newblock Cavity optomechanics: Back-action at the mesoscale.
\newblock {\em Science}, 321(5893):1172--1176, 2008.

\bibitem{doi:10.1063/1.1149175}
Malcolm~B. Gray, Daniel~A. Shaddock, Charles~C. Harb, and Hans-A. Bachor.
\newblock Photodetector designs for low-noise, broadband, and high-power
  applications.
\newblock {\em Review of Scientific Instruments}, 69(11):3755--3762, 1998.

\bibitem{Keller:90}
U.~Keller, C.~E. Soccolich, G.~Sucha, M.~N. Islam, and M.~Wegener.
\newblock Noise characterization of femtosecond color-center lasers.
\newblock {\em Opt. Lett.}, 15(17):974--976, Sep 1990.

\bibitem{refId0}
{Ir\`ene Joindot}.
\newblock Measurements of relative intensity noise (rin) in semiconductor
  lasers.
\newblock {\em J. Phys. III France}, 2(9):1591--1603, 1992.

\bibitem{Davuluri:21}
Sankar Davuluri.
\newblock Quantum optomechanics without the radiation pressure force noise.
\newblock {\em Opt. Lett.}, 46(4):904--907, Feb 2021.

\bibitem{6994803}
Lingze Duan.
\newblock Thermal noise-limited fiber-optic sensing at infrasonic frequencies.
\newblock {\em IEEE Journal of Quantum Electronics}, 51(2):1--6, 2015.

\bibitem{Hornig:22}
G.~J. Hornig, K.~G. Scheuer, E.~B. Dew, R.~Zemp, and R.~G. DeCorby.
\newblock Ultrasound sensing at thermomechanical limits with optomechanical
  buckled-dome microcavities.
\newblock {\em Opt. Express}, 30(18):33083--33096, Aug 2022.

\bibitem{yang2022micropascal}
Hao Yang, Xuening Cao, Zhi-Gang Hu, Yimeng Gao, Yuechen Lei, Min Wang, Zhanchun
  Zuo, Xiulai Xu, and Bei-Bei Li.
\newblock Micropascal-sensitivity ultrasound sensors based on optical
  microcavities.
\newblock {\em arXiv preprint arXiv:2211.07965}, 2022.

\bibitem{Krause2012}
Alexander~G. Krause, Martin Winger, Tim~D. Blasius, Qiang Lin, and Oskar
  Painter.
\newblock A high-resolution microchip optomechanical accelerometer.
\newblock {\em Nature Photonics}, 6(11):768--772, Nov 2012.

\bibitem{7435565}
Jonathan~Y. Lee and Qiang Lin.
\newblock Noise analysis in optomechanical inertial sensors.
\newblock In {\em 2016 IEEE International Symposium on Inertial Sensors and
  Systems}, pages 132--135, 2016.

\bibitem{Wissmeyer2018}
Georg Wissmeyer, Miguel~A. Pleitez, Amir Rosenthal, and Vasilis Ntziachristos.
\newblock Looking at sound: optoacoustics with all-optical ultrasound
  detection.
\newblock {\em Light: Science {\&} Applications}, 7(1):53, Aug 2018.

\bibitem{HUANG20031615}
M.~Huang.
\newblock Stress effects on the performance of optical waveguides.
\newblock {\em International Journal of Solids and Structures},
  40(7):1615--1632, 2003.

\bibitem{pierce2019acoustics}
Allan~D Pierce.
\newblock {\em Acoustics: an introduction to its physical principles and
  applications}.
\newblock Springer, 2019.

\bibitem{bedford2023introduction}
Anthony Bedford and Douglas~S Drumheller.
\newblock {\em Introduction to elastic wave propagation}.
\newblock Springer Nature, 2023.

\bibitem{Bao2018}
Gang Bao, Yixian Gao, and Peijun Li.
\newblock Time-domain analysis of an acoustic--elastic interaction problem.
\newblock {\em Archive for Rational Mechanics and Analysis}, 229(2):835--884,
  Aug 2018.

\bibitem{reed2004silicon}
Graham~T Reed and Andrew~P Knights.
\newblock {\em Silicon photonics: an introduction}.
\newblock John Wiley \& Sons, 2004.

\bibitem{narasimhamurty2012photoelastic}
Tamma~Satya Narasimhamurty.
\newblock {\em Photoelastic and electro-optic properties of crystals}.
\newblock Springer Science \& Business Media, 2012.

\bibitem{huray2009maxwell}
Paul~G Huray.
\newblock {\em Maxwell's equations}.
\newblock John Wiley \& Sons, 2009.

\bibitem{PhysRevLett.68.3603}
Xavier Gonze, Douglas~C. Allan, and Michael~P. Teter.
\newblock Dielectric tensor, effective charges, and phonons in
  \ensuremath{\alpha}-quartz by variational density-functional perturbation
  theory.
\newblock {\em Phys. Rev. Lett.}, 68:3603--3606, Jun 1992.

\bibitem{https://doi.org/10.1002/pssa.200671121}
L.~S. Hounsome, R.~Jones, M.~J. Shaw, and P.~R. Briddon.
\newblock Photoelastic constants in diamond and silicon.
\newblock {\em physica status solidi (a)}, 203(12):3088--3093, 2006.

\bibitem{Xu1992AcoustoopticD}
Jeiping Xu and Robert Stroud.
\newblock Acousto-optic devices : principles, design, and applications.
\newblock 1992.

\bibitem{6657704}
Wouter~J. Westerveld, Suzanne~M. Leinders, Pim~M. Muilwijk, Jose Pozo, Teun~C.
  van~den Dool, Martin~D. Verweij, Mirvais Yousefi, and H.~Paul Urbach.
\newblock Characterization of integrated optical strain sensors based on
  silicon waveguides.
\newblock {\em IEEE Journal of Selected Topics in Quantum Electronics},
  20(4):101--110, 2014.

\bibitem{macdonald2006noise}
David Keith~Chalmers MacDonald.
\newblock {\em Noise and fluctuations: an introduction}.
\newblock Courier Corporation, 2006.

\bibitem{H_-J_Butt_1995}
H~J Butt and M~Jaschke.
\newblock Calculation of thermal noise in atomic force microscopy.
\newblock {\em Nanotechnology}, 6(1):1, jan 1995.

\bibitem{AMABILI1996743}
M.~Amabili and M.K. Kwak.
\newblock Free vibrations of circular plates coupled with liquids: Revising the
  lamb problem.
\newblock {\em Journal of Fluids and Structures}, 10(7):743--761, 1996.

\bibitem{POMEAU2017570}
Yves Pomeau and Jarosław Piasecki.
\newblock The langevin equation.
\newblock {\em Comptes Rendus Physique}, 18(9):570--582, 2017.
\newblock Science in the making: The Comptes rendus de l’Académie des
  sciences throughout history.

\bibitem{HAUER2013181}
B.D. Hauer, C.~Doolin, K.S.D. Beach, and J.P. Davis.
\newblock A general procedure for thermomechanical calibration of
  nano/micro-mechanical resonators.
\newblock {\em Annals of Physics}, 339:181--207, 2013.

\bibitem{norton2003fundamentals}
Michael~Peter Norton and Denis~G Karczub.
\newblock {\em Fundamentals of noise and vibration analysis for engineers}.
\newblock Cambridge university press, 2003.

\bibitem{press2007numerical}
William~H Press, Saul~A Teukolsky, William~T Vetterling, and Brian~P Flannery.
\newblock {\em Numerical recipes 3rd edition: The art of scientific computing}.
\newblock Cambridge university press, 2007.

\bibitem{doi:10.1063/1.347347}
T.~R. Albrecht, P.~Grütter, D.~Horne, and D.~Rugar.
\newblock Frequency modulation detection using high‐q cantilevers for
  enhanced force microscope sensitivity.
\newblock {\em Journal of Applied Physics}, 69(2):668--673, 1991.

\bibitem{westerveld2014silicon}
Wouter~J Westerveld.
\newblock Silicon photonic micro-ring resonators to sense strain and ultrasound
  (ph. d.).
\newblock {\em Delft University of Technology}, 2014.

\end{thebibliography}
